\documentclass[prb,aps,showpacs,twocolumn]{revtex4}
\usepackage[dvipdfm]{graphicx}% Include figure files
\usepackage{dcolumn}% Align table columns on decimal point
\usepackage{bm}% bold math
\usepackage{color}
\usepackage{ulem}

\begin{document}
\newcommand{\bR}{\mbox{\boldmath $R$}}
\newcommand{\nn}{\nonumber\\}
\newcommand{\tr}[1]{\textcolor{red}{#1}}
\newcommand{\trs}[1]{\textcolor{red}{\sout{#1}}}
\newcommand{\tb}[1]{\textcolor{blue}{#1}}
\newcommand{\tg}[1]{\textcolor{green}{#1}}
\newcommand{\tbs}[1]{\textcolor{blue}{\sout{#1}}}
\definecolor{green}{rgb}{0,0.5,0.1}
\definecolor{blue}{rgb}{0,0,0.8}

\title{
Multi-orbital cluster dynamical mean-field theory with an improved continuous-time quantum Monte Carlo algorithm
}
\author{Yusuke Nomura}
\email{nomura@moegi.t.u-tokyo.ac.jp}
\author{Shiro Sakai}
\author{Ryotaro Arita}
\affiliation{Department of Applied Physics, University of Tokyo, Hongo, Bunkyo-ku, Tokyo, 113-8656, Japan.}

\date{\today}% It is always \today, today,
       
\begin{abstract}
We implement a multi-orbital cluster dynamical mean-field theory (DMFT), by improving a sample-update algorithm in the continuous-time quantum Monte Carlo method based on the interaction expansion. 
%\tb{We also develop an efficient sampling scheme for the the spin-flip and pair-hopping terms in the two-orbital case.}
The proposed sampling scheme for the spin-flip and pair-hopping interactions in the two-orbital systems mitigates the sign problem, giving an efficient way to deal with these interactions. In particular, in the single-site DMFT, 
we see that the negative signs vanish.
We apply the method to the two-dimensional two-orbital Hubbard model at half filling, where we take into account the short-range spatial correlation effects within a four-site cluster. 
We show that, compared to the single-site DMFT results, the critical interaction value for the metal-insulator transition decreases and that the effects of the spin-flip and pair-hopping terms are less significant in the parameter region we have studied.   
The present method provides a firm starting point for the study of inter-site correlations in multi-orbital systems. 
It also has a wide applicable scope in terms of realistic calculations in conjunction with density functional theory.
\end{abstract}
                            
\maketitle

\section{Introduction}
 Strongly correlated materials have attracted much interest because of their 
%fascinating diverse 
diverse fascinating
properties,~\cite{RevModPhys.70.1039} which are believed to originate from a severe competition between the 
%iteneracy
itinerancy
 and the locality of low-energy electrons.
A minimal model to describe this competition is the Hubbard model, which has been found to be surprisingly versatile despite its simple definition.
In two or three dimensions, the Hubbard model 
%\tb{usually impossible to solve analytically,
has not been solved analytically, except for several special cases,~\cite{Tasaki01041998}
%has not been solved analytically, 
and 
%hence 
therefore
we have to resort to numerical simulations.
 
The dynamical mean-field theory (DMFT),~\cite{RevModPhys.68.13} which  takes into account the dynamical local correlations accurately by mapping a lattice model onto a single impurity problem subject to a self-consistency condition,
is one of the most successful methods for describing the strong-correlation physics such as the Mott transition in infinite dimensions.~\cite{RevModPhys.68.13} 
However, the DMFT totally neglects the spatial correlations, which are essential in quantitative and also qualitative description of real materials.
For example, the single-site DMFT cannot describe the $d$-wave superconductivity observed in 
%e.g., 
high-$T_c$ cuprates. 
To overcome this problem, cluster extensions of the DMFT (cDMFT)
have been formulated.~\cite{RevModPhys.77.1027,PhysRevLett.87.186401, PhysRevB.62.R9283,PhysRevLett.91.206402}
Many studies 
on the two-dimensional (2D)
single-orbital Hubbard model 
have been performed by the cDMFT
to clarify the pseudogap phase~\cite{PhysRevLett.86.139,PhysRevB.66.075102,PhysRevB.73.165114,PhysRevB.74.125110,PhysRevB.76.045108,PhysRevLett.95.106402,PhysRevLett.97.036401,PhysRevLett.100.046402,PhysRevLett.102.056404,PhysRevB.80.045120,PhysRevB.80.165126,PhysRevB.82.045104,PhysRevB.82.134505,PhysRevB.85.035102,PhysRevLett.109.106401,SciRep.2.547,PhysRevLett.111.107001,PhysRevB.82.180511,PhysRevB.86.094522}
and the superconductivity~\cite{PhysRevLett.94.156404,PhysRevLett.95.237001,PhysRevB.74.024508,PhysRevLett.99.257002,PhysRevB.74.054513,PhysRevB.76.104509,PhysRevLett.101.156401,PhysRevB.77.184516,PhysRevB.79.195113,PhysRevB.80.172505,PhysRevB.80.205109,PhysRevLett.103.136402,PhysRevLett.108.216401,PhysRevLett.110.216405,PhysRevB.81.214525,arXiv.1308.5946} of the cuprates.

More generally, in most strongly correlated materials, 
%the bands near the Fermi level consist of several orbitals 
several orbitals are involved in the low-energy region around the Fermi level,
as exemplified 
%the transition metal oxides and the iron-based superconductors.
by the transition metal compounds and heavy fermion systems.
A description of these materials requires an extention of the Hubbard model to the multi-orbital one.
Even in the cuprates, where orbitals other than the one composing the Fermi surface are neglected in many cases, it has been proposed that the orbital degrees of freedom play a key role~\cite{PhysRevLett.105.057003,PhysRevB.85.064501,PhysRevB.87.045113} in accounting for the material dependence of the superconducting transition temperature. 

These manifest the importance of studying multi-orbital Hubbard model with including the spatial correlations. Nevertheless, it has barely been explored before because of the huge computational cost in solving the impurity problem. 
%A few exceptions are 
%\tb{the 2-site cDMFT with Hirsch-Fye algorithm~\cite{PhysRevLett.56.2521} for the three-orbital model in Ref.~\onlinecite{PhysRevLett.93.086401}},
%the 2-site cDMFT + the non-crossing approximation study for the two-orbital model in Ref.~\onlinecite{PhysRevB.79.245128}, and the 4-site cDMFT + the continuous-time quantum Monte Carlo (CTQMC)~\cite{RevModPhys.83.349,PhysRevLett.82.4155} calculation for the anisotropic two-orbital model without the spin-flip and pair-hopping terms in Ref.~\onlinecite{PhysRevLett.104.026402}.
A few exceptions are the 2-site cDMFT + the non-crossing approximation study of a two-orbital model in Ref.~\onlinecite{PhysRevB.79.245128}, 
the 2-site cDMFT + the Hirsch-Fye quantum Monte Carlo calculation~\cite{PhysRevLett.56.2521} of a three-orbital model  for Ti$_2$O$_3$ in Ref.~\onlinecite{PhysRevLett.93.086401}, and the 4-site cDMFT + the continuous-time quantum Monte Carlo (CTQMC)~\cite{RevModPhys.83.349,PhysRevLett.82.4155} calculation for an
anisotropic two-orbital model in Ref.~\onlinecite{PhysRevLett.104.026402}. In the latter two studies, the spin-flip and pair-hopping terms present in the multi-orbital Hubbard Hamiltonian were neglected. 
A study based on an accurate numerical calculation on the full multi-orbital Hamiltonian (i.e., with the spin-flip and pair-hopping terms) is still missing in literature.
Then, the aim of the present paper is to develop such a numerical scheme and to provide the first calculated results to explore the inter-site correlation physics in the multi-orbital systems.

%orbital-selective Mott transition \cite{EurPhysJB.25.191,PhysRevLett.92.216402},
%The bilayer Hubbard model without inter-layer Coulomb interactions has also been studied by cDMFT with quantum Monte Carlo solver.\cite{PhysRevB75,193103,Phys.RevB.84.180513,PhysRevB.83.045103,arXiv.1310.2821}
In the present study, we adopt the CTQMC algorithm based on the interaction expansion (CT-INT).~\cite{PhysRevB.72.035122,JETPLett.80.61}  Compared to other CTQMC algorithms,~\cite{RevModPhys.83.349} the CT-INT has an advantage in incorporating various types of interactions such as Hund's coupling and electron-phonon interacton.\cite{PhysRevB.76.035116,PhysRevB.86.155107} It also gives an efficient 
%scheme 
way
to deal with relatively large degrees of freedom, complementary to the algorithm based on the hybridization expansion,~\cite{PhysRevLett.97.076405,PhysRevB.86.155158,PhysRevB.85.115103} which is efficient for a few degrees of freedom while the computational cost grows exponentially with the degrees of freedom.
Moreover, an efficient sampling update algorithm, called submatrix update algorithm,~\cite{PhysRevB.80.195111,PhysRevB.83.075122} has recently been developed for another weak-coupling CTQMC method exploiting an auxiliary-field decomposition (CT-AUX), and has been successfully employed in cDMFT calculations 
%of 
on
the 
%two dimensional (2D)
2D~\cite{PhysRevB.85.035102,PhysRevLett.111.107001,PhysRevLett.109.106401,PhysRevLett.110.216405} and three dimensional~\cite{PhysRevB.83.075122} single-orbital Hubbard models. 
As we will show in this work, a similar submatrix update algorithm can apply to the CT-INT as well as to the multi-orbital models, too, and  it enables us to reach a strongly-correlated regime at rather low temperatures within the multi-orbital cDMFT in a reasonable computational time.
Furthermore, we develop a sampling scheme which mitigates the sign problem coming from the spin-flip and pair-hopping terms in the two-orbital 
%case.}
models.
Although in the cDMFT the negative signs remain due to the one-body hopping terms within the cluster, 
in the single-site DMFT, we see that the proposed method completely eliminates the negative signs.
%the proposed scheme completely resolves the sign problem in the single-site DMFT.

We apply the method to the 2D two-orbital Hubbard model on a square lattice within the 4-site cellular DMFT.\cite{PhysRevLett.87.186401}
%method
We show that the short-range spatial correlations reduce the critical interaction strength of the Mott metal-insulator transition substantially. 
We also find that the model with the Ising-type Hund's coupling overestimates the tendency toward the insulating phase while the difference between the results with and without the spin-flip and pair-hopping terms is less significant than that of the single-site DMFT.

This paper is organized as follows. In Sec.~\ref{sec_method}, we briefly review the CT-INT algorithm and show how the submatrix update 
and the efficient update scheme for the non-density-density interactions are
%is 
incorporated into the algorithm.
We show the cellular DMFT results for the 2D two-orbital Hubbard model in Sec.~\ref{sec_result}.
Section~\ref{sec_sum} is devoted to the conclusion. 
The derivation of the several equations used in Sec.~\ref{sec_method}, and
a proof of the absence of negative signs in the two-orbital models in our scheme
% and 
%supplementary self-energy data for the 2D two-orbital Hubbard model
are given in Appendices.
  
\section{Method}\label{sec_method}
In this section, we explain, in detail, the schemes employed in our calculations. 
 Sec.~\ref{sec_expansion} and Sec.~\ref{sec_Monte} are devoted to a brief introduction of the CT-INT algorithm.
 Sec.~\ref{sec_submatrix} shows how the submatrix update scheme, which has been employed only in the Hirsch-Fye and CT-AUX algorithms in literature, is incorporated in the CT-INT method.
In Sec.~\ref{subsec_multi}, we show the extension to the single-site multi-orbital case, where we propose an efficient 
sampling scheme for the spin-flip and pair-hopping terms, double-vertex update, in the two-orbital case.
Finally, we show the extension to multi-site multi-orbital case in Sec.~\ref{sec_mo_ms}.
 
\subsection{Single-orbital case}
\subsubsection{Interaction expansion of partition function}\label{sec_expansion}
%Rubtsov {\it et al.} \cite{JETPLett.80.61,PhysRevB.72.035122} developed the CT-INT algorithm, which has been used for solving the impurity problem in the DMFT.
The CT-INT algorithm was developed by Rubtsov {\it et al.} \cite{JETPLett.80.61,PhysRevB.72.035122}
Here we review the basic part of the algorithm in order to define our notations used in the next section.
We first consider the single-orbital and single-impurity model for simplicity.

The action for the single-orbital impurity problem reads
\begin{eqnarray}
 S_{\rm{imp}} =  S_0 + S_{\rm{int}},
\end{eqnarray}
where
\begin{eqnarray}
S_0 = - \int^{\beta}_{0} d \tau  \int^{\beta}_{0} d \tau'  \sum_{\sigma}
\mathcal{G}_{0\sigma} ^{-1} (\tau - \tau') \hat{c}^{\dagger}_{\sigma}(\tau) \hat{c}^{\ }_{\sigma} (\tau')
\end{eqnarray}
and 
\begin{eqnarray}
 S_{\rm{int}}
= \int^{\beta}_{0}  d \tau 
 U \hat{n}_{ \uparrow} (\tau) \hat{n}_{ \downarrow} (\tau)
\end{eqnarray}
with the inverse temperature $\beta$, the bath Green's function $\mathcal{G}_{0\sigma}$, and the Hubbard interaction $U$.
$\hat{c}_{\sigma}^{\dagger}$ ($\hat{c}_\sigma$) is a Grassmann variable representing the creation (annihilation) of an impurity electron with the spin $\sigma$, and $\hat{n}_{\sigma} = \hat{c}_{\sigma}^{\dagger} \hat{c}_{\sigma}$. 

In order to reduce the sign problem, we introduce additional parameters $\alpha_\sigma$ defined as~\cite{PhysRevB.76.035116} 
\begin{eqnarray}
%&&\alpha_{\uparrow}(s=\pm1) = 1/2 + s \delta,  \nn
%&&\alpha_{\downarrow}(s=\pm1) = 1/2 - s  \delta   
&&\alpha_{\uparrow}(s) = 1/2 + s \delta,  \nn
&&\alpha_{\downarrow}(s) = 1/2 - s  \delta
\label{def_alpha}
\end{eqnarray} 
with $\delta = 1/2 + 0^+ $ and $s=\pm1$. In practice,  we typically set $0^+$ to be the order of $10^{-2}$.
In the absence of this $\alpha$ term, we suffer from the negative sign problem because the elements of the $V$ matrix 
corresponding to the $U$ vertex
in Eq.~(\ref{Eq_werner_wt}) can
take negative values.~\cite{Werner_private_comm}
Then 
%we rewrite the action as 
the action is recast into
\begin{eqnarray}
S_0 = - \int^{\beta}_{0} d \tau  \int^{\beta}_{0} d \tau'  \sum_{\sigma}
\tilde{\mathcal{G}}_{0\sigma} ^{-1} (\tau - \tau') \hat{c}^{\dagger}_{\sigma}(\tau) \hat{c}^{\ }_{\sigma} (\tau')
\end{eqnarray}
and 
\begin{eqnarray}
 S_{\rm{int}}
= \int^{\beta}_{0}  d \tau 
\sum_{s = \pm 1 } \frac{U}{2} 
 \bigl[ \hat{n}_{\uparrow} (\tau) -  \alpha_{\uparrow}(s) \bigr ] 
 \bigl[ \hat{n}_{\downarrow} (\tau) -  \alpha_{\downarrow}(s) \bigr ],  
\end{eqnarray}
where $\tilde{\mathcal{G}}_{0\sigma}$ is the Weiss function defined with a new chemical potential 
$\tilde{\mu} = \mu - U/2$.
%By performing the perturbation expansion with respect to $S_{\rm int}$, we write the partition function as 
The perturbation expansion with respect to $U$ term leads to
\begin{widetext}
\begin{eqnarray}
\frac{Z}{Z_0} &=& 
\sum_{n=0}^{\infty}  \left(- \frac{U}{2} \right)^n \int_{0}^{\beta} d \tau_1 \sum_{s_1=\pm1}
\cdots 
\int_{0}^{\tau_{n\!-\!1}} d \tau_{n} \sum_{s_{n} =\pm 1}
\prod_{\sigma} \bigl \langle  
\bigl [   \hat{n}_{\sigma} (\tau_1) -  \alpha_{\sigma}(s_1)  \bigr ]  \cdots 
\bigl [   \hat{n}_{\sigma} (\tau_n) -  \alpha_{\sigma}(s_n)  \bigr ] 
 \bigr \rangle_0
\nn 
&=& 
\sum_{n=0}^{\infty}  \left(- \frac{U}{2} \right)^n \int_{0}^{\beta} d \tau_1 \sum_{s_1=\pm1}
\cdots \int_{0}^{\tau_{n\!-\!1}} d \tau_{n} \sum_{s_{n}=\pm1}
\prod_{\sigma}  {\rm det}  A'_{\sigma}(\{ s_i, \tau_i \})
\label{Eq.Z}
\end{eqnarray}
\end{widetext}
where 
%\begin{eqnarray}
$Z_0 = \int \mathcal{D} [ \hat{c}^{\dagger},  \hat{c} ] e^{-S_0[\hat{c}^{\dagger},  \hat{c}]}$
%\end{eqnarray}
is a noninteracting partition function
and the thermal average for the products of Grassmann variables  
$\langle V[ \hat{c}^{\dagger},  \hat{c} ] \rangle_0$ is defined as  
\begin{eqnarray}
\bigl \langle  V[ \hat{c}^{\dagger},  \hat{c} ]  \bigr \rangle_0 = \int \mathcal{D} [ \hat{c}^{\dagger},  \hat{c} ]     e^{-S_0[\hat{c}^{\dagger},  \hat{c}]} V [ \hat{c}^{\dagger},  \hat{c} ].
\end{eqnarray}
$A'_{\sigma} (\{ s_i, \tau_i \})$ is an $n\times n$ matrix whose element is given by 
\begin{eqnarray}
\bigl[A'_{\sigma} (\{ s_i, \tau_i \}) \bigr ]_{ij} = \tilde{\mathcal{G}}_{0\sigma}(\tau_i - \tau_j) -  \alpha_{\sigma}(s_i)  \delta_{ij}.
\end{eqnarray}
With a function
\begin{eqnarray}
 f_{\sigma}(s) 
= \left \{
 \begin{array}{ll} 
    \frac{\alpha_{\sigma} (s)} {\alpha_{\sigma} (s)- 1 }   &  s =\pm1 \\
    1  & s = 0 
 \end{array}
 \right . , 
\end{eqnarray}
a configuration 
\begin{eqnarray}
C_n = \{ (s_1 , \tau_1),  \cdots , (s_n, \tau_n)  \}, 
\end{eqnarray}
%and a notation for the ``{\it sum}'' over the configuration space, 
%\begin{eqnarray}
%\sum_{C_n} = \sum_{n=0}^{\infty}   \int_{0}^{\beta}   \sum_{s_1=\pm1}
%\cdots \int_{0}^{\tau_{n\!-\!1}}  \sum_{s_n=\pm1}, 
%\cdots \int_{0}^{\beta}  \sum_{s_n=\pm1}, 
%\end{eqnarray}
%\begin{eqnarray}
%\sum_{C_n} = \sum_{n=0}^{\infty}  \frac{1}{n !} \int_{0}^{\beta}   \sum_{s_1=\pm1}
%\cdots \int_{0}^{\tau_{n\!-\!1}}  \sum_{s_n=\pm1}, 
%\cdots \int_{0}^{\beta}  \sum_{s_n=\pm1}, 
%\end{eqnarray}
Eq.~(\ref{Eq.Z}) is rewritten as 
\begin{eqnarray}
\frac{Z}{Z_0} &=& 
\sum_{n=0}^{\infty}   \int_{0}^{\beta}   \sum_{s_1=\pm1}
\cdots \int_{0}^{\tau_{n\!-\!1}}  \sum_{s_n=\pm1}  \nn
&&\left[ \prod_{i=1}^{n} \frac{K(s_i) d \tau_i}{2 \beta}  \times
  \prod_{\sigma}  {\rm det}  A_{\sigma}(C_n) \right],   
  \label{Eq.ZA}
\end{eqnarray}
where
\begin{eqnarray}
&&K(s) =  \frac{-\beta U}{ ( f_{\uparrow} (s) -1 ) (f_{\downarrow} (s) - 1 ) }\ {\rm for}\ s=\pm 1, 
\nn 
&& A_{\sigma} (C_n)  =   F_{\sigma}^{\{s_i\}} -  G_{0\sigma}^{\{\tau_i\}} (F_{\sigma}^{\{s_i\}}-1) .
\label{Eq.A}
\end{eqnarray} 
Here,  we define $n\times n$ matrices $G_{0\sigma}^{\{\tau_i\}}$ and $F_{\sigma}^{\{s_i\}}$, whose elements are 
\begin{eqnarray}
\Bigl[G_{0\sigma}^{\{\tau_i\}} \Bigr]_{ij} = \tilde{\mathcal{G}}_{0\sigma}(\tau_i - \tau_j) 
\end{eqnarray}
 and 
 \begin{eqnarray}
\Bigl[F_{\sigma}^{\{ s_i \}} \Bigr]_{ij} =  f_{\sigma} (s_i) \delta_{ij},  
\end{eqnarray}
respectively. 
Since the equality $K(s=1) = K(s=-1)$ holds for our choice of $\alpha_{\sigma}$ (Eq.(\ref{def_alpha})), we will simply denote them as $K$ hereafter.

\subsubsection{Monte Carlo sampling}\label{sec_Monte}
According to Eq.~(\ref{Eq.ZA}), the weight for the configuration $C_n$ is given by 
\begin{eqnarray}
W(C_n) = \left ( \frac{K d\tau}{2 \beta}  \right ) ^n  \times  \prod_{\sigma}  {\rm det}  A_{\sigma}(C_n). 
\end{eqnarray}
To guarantee the ergodicity, the addition and removal of the vertices with a random orientation of the auxiliary Ising spins $s_i = \pm1$ at randomly-chosen imaginary times $ \tau_i \in [0,\beta)$ are sufficient. 
To add a vertex, we randomly pick an imaginary time from the range $[0,\beta)$ and put there an auxiliary Ising spin with a 
randomly-chosen orientation, with a proposal probability of $P_{0}(C_n \rightarrow C_{n+1}) = d \tau /2\beta $. 
To remove a vertex, we randomly choose one of the existing vertices, 
with the proposal probability
$P_{0}(C_{n+1} \rightarrow C_n) = 1 / (n+1) $. 
In the Metropolis algorithm, the acceptance ratio is 
%given by 
\begin{eqnarray}
P(C \rightarrow C') = {\rm min} \!  \left( \frac{ W (C') P_0 (C' \rightarrow C )} { W(C) P_0 (C \rightarrow C')} , 1   \right). 
\end{eqnarray}
Applying this to the CT-INT, we obtain the acceptance ratios 
\begin{eqnarray}
P(C_n \! \rightarrow \! C_{n+1}) = {\rm min} \left( \frac{K}{n+1} 
\prod_{\sigma} \frac{ {\rm det} A_{\sigma}(C_{n+1})}{{\rm det} A_{\sigma}(C_n)}, 1   \right)
\label{Eq.add}
\end{eqnarray}
for the addition of a vertex, and 
\begin{eqnarray}
P(C_{n+1} \! \rightarrow \! C_{n}) = {\rm min} \!  \left( \frac{n+1} {K}
\prod_{\sigma} \frac{{\rm det} A_{\sigma}(C_n)}{ {\rm det} A_{\sigma}(C_{n+1})} , 1   \right)
\label{Eq.rem}
\end{eqnarray} 
for the removal of a vertex.

\subsubsection{Submatrix update}\label{sec_submatrix}
In the conventional fast update scheme, the matrix $A_\sigma^{-1}$ is updated at each change of the auxiliary spins.
Nukala {\it et al.} \cite{PhysRevB.80.195111} and subsequently Gull {\it et al.} \cite{PhysRevB.83.075122} introduced a more efficient update algorithm, called submatrix update, to the Hirsch-Fye and the CT-AUX quantum Monte Carlo algorithms, respectively, where the matrix $A_\sigma^{-1}$ is updated at once after $k_{\rm max}$-time updates are done.
The speed-up comes not from the reduction of the operation times, but from
%but from employing the matrix operation instead of vector one to fit the modern computer architectures, which 
an efficient memory management by employing the matrix (submatrix) which is accommodated in a cache memory of the modern computer architectures, as
is detailed in Ref~\onlinecite{PhysRevB.83.075122}.
Here we introduce a similar submatrix update algorithm to the CT-INT,
which is essential for implementing the multi-orbital cDMFT calculation, described in Sec.~\ref{subsec_multi}, in a practical computational time. We refer the readers to Refs.~\onlinecite{PhysRevB.80.195111,PhysRevB.83.075122} for a detailed derivation of Eqs.~(\ref{Eq.accratio1}), (\ref{Eq.accratio2}), and (\ref{Eq.Aupdate}) below, for which we avoid a repetition.

In the following we omit the spin index $\sigma$ for simplicity while the procedure described below has to be done for both spins $\sigma = \uparrow$ and $\downarrow$. We start from a configuration $C^0_n$.  Suppose we know the corresponding matrix $A_{0}^{-1} (C^0_{n})$ and that we propose 
%the 
insertions or removals of the auxiliary spins (vertices) for the next $k_{\rm max}$ times; let $k_{\rm max}^{\rm ins}$ be the number of the insertions. We define an extended configuration $\tilde{C}^0_{n+k_{\rm max}^{\rm ins}}$, which is comprised of the original configuration $C^0_n$ and the $k_{\rm max}^{\rm ins}$ ``{\it noninteracting}" vertices added at randomly-chosen imaginary times, i.e., 
\begin{eqnarray}
\tilde{C}^0_{n+k_{\rm max}^{\rm ins}} = \{ \underbrace{ (s_1^0 , \tau_1^0),  \cdots , (s_n^0, \tau_n^0)}_{\text{\Large$C^0_n$}} ,  (s^0_{n+1} = 0  , \tau^0_{n+1}), \nn \cdots, (s^0_{n+k_{\rm max}^{\rm ins}} = 0, \tau^0_{n+k_{\rm max}^{\rm ins}})\}.   
\end{eqnarray}

Then, we accordingly define an extended $(n+k_{\rm max}^{\rm ins}) \times (n+k_{\rm max}^{\rm ins})$ matrix $\tilde{A}_{0}^{-1}(\tilde{C}^0_{n+k_{\rm max}^{\rm ins}})$ by 
\begin{eqnarray}
\tilde{A}_{0}^{-1} = 
%(\tilde{C}^0_{n+k_{\rm ins}})  
\left( \begin{array} {ll}
A_0  & 0  \\ 
B & 1 \\ 
\end{array} 
\right )^{-1}
 = 
 \left( \begin{array} {cc}
A_0^{-1}  & 0  \\ 
-B A_0^{-1}& 1 \\ 
\end{array} 
\right ).
\end{eqnarray}
Here, $B$ is a $k_{\rm max}^{\rm ins} \times n$ matrix with elements $B_{ij} = - \tilde{\mathcal{G}}_0 (\tau^0_{n+i} - \tau^0_j) (f(s^0_j) - 1)$. Notice that the equality ${\rm det} A_{0}(C_n) = {\rm det} \tilde{A}_{0}(\tilde{C}_{n+k_{\rm max}^{\rm ins}})$ holds, which is utilized in the calculation of the acceptance ratio described below.

With the extended matrix $\tilde{A}_{0}^{-1}$ and configuration $\tilde{C}^0_{n+k_{\rm max}^{\rm ins}}$, the addition and the removal of the vertices can be done by just flipping the orientation of the auxiliary spins: The addition is expressed by changing an auxiliary spin $s$ from 0 to $\pm 1$ while the removal is expressed by the change from $\pm 1$ to 0. Since the number of auxiliary spins (including those with zero value) is fixed during the spin-flip process, we abbreviate $\tilde{C}^0_{n+k_{\rm max}^{\rm ins}}$ to $\tilde{C}^0$ below.

For later use, we denote the configuration after $k(<k_{\rm max})$-th
 updates by $\tilde{C}^k$ and the auxiliary spins in $\tilde{C}^k$ by $ \{ s^k_{i} \} $.
% $1 \leq i \leq n+k_{\rm max}^{\rm ins}$. 
The positions
%(in 1 to $n+k_{\rm max}^{\rm ins}$)
 of the flipped spins are denoted by $p_j$ ($j= 1,2, \cdots, l_k$; $ 1 \le p_j \le n+k_{\rm max}^{\rm ins}$) with $l_k$ being the number of the flipped spins. 
With these notations, we define an $l_k \times l_k$ matrix $\Gamma_{k}$ by 
\begin{eqnarray}
\bigl[\Gamma_k \bigr]_{ij} =  \bigl[ \tilde{G}(\tilde{C}^0) \bigr]_{p_i p_j} - \delta_{ij} \frac{1 + \gamma(s^k_{p_i}, s^0_{p_i})}{\gamma(s^k_{p_i}, s^0_{p_i})}, 
\label{Eq.Gam}
\end{eqnarray}
with
\begin{eqnarray}
\gamma(s',s) = \frac{ f(s') - f(s) } {f(s)}.
\label{Eq.gamfunc}
\end{eqnarray}
The elements of the Green's function matrix $ [\tilde{G}(\tilde{C}^0)]_{ij}$ can be efficiently calculated by using Eq.~(\ref{Eq.Ge}) for $1 \leq j \leq n$. 
For $n+1 \leq j \leq n+k_{\rm max}^{\rm ins}$, we need to use Eq.~(\ref{Eq.Gb}) to compute them since $s_j = 0$. 
The matrix $\Gamma_k^{-1}$ is updated at each change of the auxiliary spins and is used to calculate the acceptance ratio. An efficient method to update $\Gamma_k^{-1}$ is elaborated in Ref.~\onlinecite{PhysRevB.83.075122} and we do not repeat it here.

The acceptance ratios, Eqs.~(\ref{Eq.add}) and (\ref{Eq.rem}), can also be calculated easily from $\Gamma_k^{-1}$. 
Let us consider a $(k+1)$-th update at which the $p$-th spin is proposed to change from $s^k_{p}$ to $s'^k_{p}$ and the configuration moves from $\tilde{C}^{k}$ to $\tilde{C}'^{k}$. 
When $p\neq p_j$ for $j=1,2,\cdots l_k$, the determinant ratio is given by
\begin{eqnarray}
\frac{{\rm det} \tilde{A}'_k }{ {\rm det} \tilde{A}_k }  =  -  \gamma(s'^k_p, s^k_p)\frac{{\rm det} \Gamma'_{k} }{  {\rm det} \Gamma_k},
\label{Eq.accratio1}
\end{eqnarray}
where $\Gamma'_{k}$ is an $(l_k+1) \times (l_k+1)$ matrix whose elements of the $(l_k+1)$-th row and column are calculated from Eq.~(\ref{Eq.Gam}) with $p_{l_k+1} = p$.
Otherwise, $p$ coincides with one of $\{p_j\} (j=1,2,\cdots l_k)$, i.e., a previously inserted vertex is proposed to be removed. 
In this case, the $p$-th spin has already been changed from $s^0_p =0$ to $s^k_p = \pm 1$, and therefore $s'^k_p = 0 = s^0_p$. Then the determinant ratio is given by
\begin{eqnarray}
\frac{{\rm det} \tilde{A}'_k }{ {\rm det} \tilde{A}_k }  =  -  \frac{1}{\gamma(s^k_p,0)} \frac{{\rm det} \Gamma'_{k} }{  {\rm det} \Gamma_{k}}. 
\label{Eq.accratio2}
\end{eqnarray}
Here $\Gamma'_{k}$ is an $(l_k-1) \times (l_k-1)$ matrix in which a column and a row corresponding to $p$-th spin are removed from $\Gamma_{k}$.

If the proposal is accepted, the proposed configuration $\tilde{C}'^{k}$ becomes the new configuration $\tilde{C}^{k+1}$, and accordingly, the size of the $\Gamma$ matrix increases or decreases.   
Otherwise, the configuration and the $\Gamma$ matrix are unchanged.
Then, we move to the $(k+2)$-th update.
We repeat this procedure up to $k_{\rm max}$ times.  

After $k_{\rm max}$-th update, we recompute the $A^{-1}$ matrix.  
To this end, we use the identity~\cite{PhysRevB.80.195111,PhysRevB.83.075122}
\begin{eqnarray}
\label{Eq.Aupdate}
\bigl[ \tilde{A}_{k_{\rm max}}^{-1} \bigr ]_{ij}  = 
 \frac{ \bigl[\tilde{A}_0^{-1} \bigr]_{ij} - \bigl [\tilde{G}(\tilde{C}^0) \bigr]_{i p_k} \bigl[ \Gamma_{k_{\rm max}}^{-1} \bigr]_{p_k p_l}  \bigl[\tilde{A}_0^{-1} \bigr]_{p_l j} }
 {1+\gamma(s^{k_{\rm max}}_i,s^0_i)}. \nn 
\end{eqnarray} 
We then delete the ``{\it noninteracting}" auxiliary spins from $\tilde{A}^{-1}_{k_{\rm max}}$ by removing the corresponding rows and columns and obtain a new $A^{-1}$ matrix, which gives the starting point for the next $k_{\rm max}$-times updates.

\subsection{Multi-orbital case}\label{subsec_multi}
%\subsubsection{\tb{Straightforward extension of the single-orbital method to the multi-orbital case}}
\subsubsection{Extension to the multi-orbital systems with the conventional single-vertex update}\label{sec_mo_single}
We now extend the above algorithm to the multi-orbital case.
The action of the multi-orbital impurity problem is given by 
\begin{eqnarray}
 S_{\rm{imp}} =  S_0 + S_{\rm{int}},
\end{eqnarray}
where
\begin{eqnarray}
\ \ S_0 = - \int^{\beta}_{0} d \tau  \int^{\beta}_{0} d \tau'  \sum_{ij,\sigma}%\sum_{\sigma}
\bigl [ \mathcal{G} ^{-1}_{0\sigma} (\tau - \tau') \bigr]_{ij} \hat{c}^{\dagger}_{i\sigma}(\tau) \hat{c}^{\ }_{j\sigma} (\tau')
\nn
\end{eqnarray}
and 
\begin{eqnarray}
  \ S_{\rm{int}}
= \int^{\beta}_{0} \! d \tau  \! \! \! \! 
&\biggl[&  \! \! \! 
\sum_{i} U \hat{n}_{i \uparrow} (\tau) \hat{n}_{i \downarrow} (\tau)
+ 
%\int^{\beta}_{0} d \tau 
 \sum_{i<j,\sigma}%\sum_{\sigma}
U' \hat{n}_{i\sigma} (\tau) \hat{n}_{j \overline{\sigma}} (\tau)
\nn
&+& 
% \int^{\beta}_{0} d \tau 
 \sum_{i<j,\sigma}%\sum_{\sigma}
(U'-J_{\rm H}) \hat{n}_{ i \sigma} (\tau) \hat{n}_{j \sigma} (\tau)
\nn
&+& 
% \int^{\beta}_{0} d \tau 
 \sum_{i \neq j} 
J_{\rm H}
\hat{c}^{\dagger}_{i\uparrow} (\tau)
\hat{c}^{\ }_{j\uparrow} (\tau)
\hat{c}^{\dagger}_{j\downarrow}(\tau)
\hat{c}^{\ }_{i\downarrow}(\tau)
\nn
&+& 
% \int^{\beta}_{0} d \tau 
 \sum_{i \neq j} 
J_{\rm H}
\hat{c}^{\dagger}_{i\uparrow}(\tau)
\hat{c}^{\ }_{j\uparrow} (\tau)
\hat{c}^{\dagger}_{i\downarrow}(\tau)
\hat{c}^{\ }_{j\downarrow}(\tau) \biggr]. 
\end{eqnarray}
Here, the Weiss function $ \mathcal{G} ^{-1}_{0\sigma} (\tau - \tau')$
is a matrix with respect to the orbital $i$ and $j$.
 $U$, $U'$, and $J_{\rm H}$ are the intra-orbital Coulomb interaction, inter-orbital Coulomb interaction, and Hund's coupling, respectively.  
$\hat{c}_{i\sigma}^{\dagger}$ ($\hat{c}_{i\sigma}$) is a Grassmann variable representing the creation (annihilation) of the impurity electron with the orbital $i$ and the spin $\sigma$, and $\hat{n}_{i\sigma} = \hat{c}_{i\sigma}^{\dagger} \hat{c}_{i\sigma}$. 

As in the single-orbital case, we introduce additional parameters. We employ~\cite{PhysRevB.80.155132} 
\begin{eqnarray}
%\alpha_{1\uparrow}(s=\pm1) = 1/2 + s \delta_1  \nn
%\alpha_{1\downarrow}(s=\pm1) = 1/2 - s  \delta_1   
\alpha_{1\uparrow}(s) = 1/2 + s \delta_1  \nn
\alpha_{1\downarrow}(s) = 1/2 - s  \delta_1   
\end{eqnarray}
with $s=\pm 1$ and $\delta_1 = 1/2 + 0^+ $, and 
\begin{eqnarray}
%\alpha_{2\uparrow}(s=\pm1) &=& + s \delta_2 \nn
%\alpha_{2\downarrow}(s=\pm1) &=&  - s \delta_2    
\alpha_{2\uparrow}(s) &=& + s \delta_2 \nn
\alpha_{2\downarrow}(s) &=&  - s \delta_2
\end{eqnarray}
with a small positive real number $\delta_2$.
Then we rewrite the 
%effective 
non-interacting part of the
action as 
\begin{eqnarray}
\ \ S_0 =  - \int^{\beta}_{0} d \tau  \int^{\beta}_{0} d \tau'  \sum_{ij,\sigma} 
\bigl[ \tilde{\mathcal{G}} ^{-1}_{0\sigma} (\tau - \tau') \bigr]_{ij} 
 \hat{c}^{\dagger}_{i\sigma}(\tau) \hat{c}^{\ }_{j\sigma} (\tau'), 
\nn 
\label{Eq.S0_multi}
\end{eqnarray}
where $\tilde{\mathcal{G}}_{0\sigma}$ is the local noninteracting Green's function defined at a modified chemical potential 
$\tilde{\mu} = \mu - U/2 -  N_{\rm {orb}} (2U'-J_{\rm H}) /2 $ with $N_{\rm {orb}}$ being the number of the orbitals.
The interaction part of the action is 
\begin{widetext}
\begin{eqnarray}
 S_{\rm{int}}    &=&  \int^{\beta}_{0} d \tau  \sum_{s=\pm1} \Biggl[
 \sum_{i}  \frac{U}{2} 
 \bigl[ \hat{n}_{i \uparrow} (\tau) -  \alpha_{1\uparrow}(s) \bigr ] 
 \bigl[ \hat{n}_{i \downarrow} (\tau) -  \alpha_{1\downarrow}(s) \bigr ] 
+     \sum_{i<j, \sigma}%\sum_{\sigma}
\frac{U'}{2} 
 \bigl[ \hat{n}_{i\sigma} (\tau) -  \alpha_{1\sigma}(s) \bigr ]  
 \bigl[ \hat{n}_{j \overline{\sigma}} (\tau) -  \alpha_{1\overline{\sigma}}(s) \bigr ] 
\nn
&+&     \sum_{i<j,\sigma}%\sum_{\sigma}
\frac{U'-J_{\rm H} } {2}
 \bigl[ \hat{n}_{i\sigma} (\tau) -  \alpha_{1\sigma}(s) \bigr ]  \bigl[ \hat{n}_{j \sigma} (\tau) -  \alpha_{1\sigma}(s) \bigr ] 
+    \sum_{i \neq j} \frac{J_{\rm H}}{2}
\bigl[ \hat{c}^{\dagger}_{i\uparrow} (\tau) \hat{c}^{\ }_{j\uparrow} (\tau) - \alpha_{2\uparrow}(s) \bigr ]  
\bigl[ \hat{c}^{\dagger}_{j\downarrow}(\tau) \hat{c}^{\ }_{i\downarrow}(\tau)  - \alpha_{2\downarrow}(s) \bigr ] 
\nn
&+&   \sum_{i \neq j} \frac{J_{\rm H}}{2}
\bigl[ \hat{c}^{\dagger}_{i\uparrow}(\tau) \hat{c}^{\ }_{j\uparrow} (\tau)- \alpha_{2\uparrow}(s) \bigr ] 
\bigl[ \hat{c}^{\dagger}_{i\downarrow}(\tau) \hat{c}^{\ }_{j\downarrow}(\tau)  - \alpha_{2\downarrow}(s) \bigr ] 
\Biggr].
\label{Eq.Sint}
\end{eqnarray}
\end{widetext}
Thanks to the $\alpha_{1}$ terms, we can avoid the negative 
signs coming from the density-density interactions 
as in the Hirsch-Fye and CT-AUX algorithms.~\cite{PhysRevB.80.155132} 
Without them, negative signs appear since the $V$ matrix 
corresponding to the density-density-type vertex
in Eq.~(\ref{Eq_werner_wt})
obtains matrix elements with negative values.~\cite{Werner_private_comm}
On the other hand, the number of negative signs increases with $\delta_2$.
However, 
%in the case where 
as far as
the off-diagonal parts of the Weiss function vanish 
(i.e., $[\tilde{\mathcal{G}}_{0\sigma} ]_{ij} = 0$ for $i\neq j$), we need a non-zero $\delta_2$ value to satisfy the ergodicity. 
In the two-orbital case, we can incorporate the last two terms in Eq.~(\ref{Eq.Sint}) more efficiently, as we shall discuss in Sec.~\ref{sec_dbup}.
%($\sim 10^{-4}$) for a stable calculation~\cite{PhysRevB.80.155132}.

If we neglect the spin-flip and pair-hopping terms, which correspond to the last two terms in Eq.~(\ref{Eq.Sint}), we only have the density-density type interactions and the symmetry of the spin lowers from SU(2) to $Z_2$.    
This mitigates the sign problem considerably and hence often employed in literature
though the neglect has no physical ground.~\cite{EurPhysJB.5.473,PhysRevB.70.172504,PhysRevB.74.155102,PhysRevB.70.054513}
Hereafter, we call the Hamiltonian with the spin-flip and pair-hopping terms as SU(2)-symmetric Hamiltonian, and the Hamiltonian without them as $Z_2$-symmetric Hamiltonian.  

In the multi-orbital case, we define a configuration as 
\begin{eqnarray}
 C_n = \{ (\kappa_1,s_1,\tau_1), \cdots, (\kappa_n,s_n,\tau_n)\},
\end{eqnarray}
where we introduce the index $\kappa$ for the type of the interaction. 
%which runs from 1 to $4N_{\rm orb}^2 - 3 N_{\rm orb}$ for SU(2)-symmetric case, and from 1 to $2N_{\rm orb}^2 - N_{\rm orb}$ for $Z_2$-symmetric case. 
%A ``sum" over the configuration space is now
%\begin{eqnarray}
%\ \ \sum_{C_n} = \sum_{n=0}^{\infty}  \int_{0}^{\beta}   \sum_{\kappa_1}\sum_{s_1=\pm1}
%\cdots \int_{0}^{\tau_{n\!-\!1}}  \sum_{\kappa_n}  \sum_{s_n=\pm1}.  \nn
%\end{eqnarray}
We also need to generalize the $f$ and $K$ functions:  
In the case where $\kappa$ designates a density-density interaction, we define $f$ as 
\begin{eqnarray}
 f_{\kappa\sigma}(s) = \left \{
 \begin{array}{ll} 
    \frac{\alpha_{1\sigma} (s)} {\alpha_{1\sigma} (s)- 1 }   &  s =\pm1 \\
    1  & s = 0 
 \end{array}
 \right . , 
\end{eqnarray}
otherwise, it is defined as 
\begin{eqnarray}
 f_{\kappa\sigma}(s) = \left \{
 \begin{array}{ll} 
    \frac{\alpha_{2\sigma} (s)} {\alpha_{2\sigma} (s)- 1 }   &  s =\pm1 \\
    1  & s = 0 
 \end{array}
 \right . . 
\end{eqnarray}
Then the $K$ function is defined by
\begin{eqnarray}
K_{\kappa}(s) =  \frac{-\beta V_{\kappa} }{ ( f_{\kappa\uparrow} (s) -1 ) (f_{\kappa\downarrow} (s) - 1 ) }, 
\end{eqnarray}
for $s=\pm 1$ with $V_{\kappa}= U, U', U'-J_{\rm H}, {\rm or} \ J_{\rm H}$. 

With these functions, the partition function for the multi-orbital impurity problem is written in the form
\begin{eqnarray}
\frac{Z}{Z_0} &=& 
\sum_{n=0}^{\infty}  \int_{0}^{\beta}   \sum_{\kappa_1}\sum_{s_1=\pm1}
\cdots \int_{0}^{\tau_{n\!-\!1}}  \sum_{\kappa_n}  \sum_{s_n=\pm1} \nn
 &&\left[ \prod_{i=1}^{n} \frac{K_{\kappa_i}(s_i) d\tau_i}{2 \beta}  \times
  \prod_{\sigma}  {\rm det}  A_{\sigma}(C_n) \right]. 
  \label{Eq.Z_multi}
\end{eqnarray}
The $A$ matrix has a similar form as that in Eq.~(\ref{Eq.A}), but now we have an additional orbital indices for the $G_{0}$ matrix and $\kappa$ index for the $F$ matrix. 
When the interaction between the same spin (the third term in Eq.~(\ref{Eq.Sint})) is inserted, the size of the $A$ matrix for that spin increases by two, while no increase for the opposite spin. 
Therefore, the size of the $A$ matrix does not necessarily agree with the number of the interaction vertices $n$, while (size of $A_{\uparrow}$) + (size of $A_{\downarrow}$) = $2n$ holds.

Now the application of the submatrix update to the multi-orbital case is straightforward. 
We only comment on several important 
%points.
differences from the single-orbital one.
 (i) We need  to modify the definition of the $\gamma$ function to have $\kappa$ index. 
(ii) As in the $A$ matrix, the sizes of the $\Gamma_{\uparrow}$ and $\Gamma_{\downarrow}$ matrices do not necessarily agree. 
(iii) If the update is related to the interaction between the same spin, we need to enlarge or shrink the $\Gamma$ matrix by two rows and two columns only for the relevant spin components. 

\subsubsection{Efficient sampling scheme for the spin-flip and pair-hopping terms:
% Two orbital case
Double-vertex update}\label{sec_dbup}
%The above algorithm concentrate on the single-vertex-insertion or single-vertex-removal updates.
%In this section, we will show that, as for the spin-flip and pair-hopping terms in the two-orbital model, we only need to insert or remove two vertices simultaneously.  
%The double-vertex insertion or double-vertex removal are shown to be more efficient than single-vertex updates, 
%especially in the sense that they mitigate the sign problem. 
%Hereafter, in this section, we focus on the two-orbital case.

Here, 
%in the two-orbital Hubbard model without the hybridization between the orbitals, we show 
we show, in the two-orbital Hubbard model without a hybridization between the orbitals,
that the spin-flip and pair-hopping interactions can be treated efficiently by incorporating the double-vertex insertion and removal processes, on top of the standard single-vertex updates for the density-density-type interactions. 
The double-vertex update allows the spin-flip and pair-hopping interactions to appear only at even perturbation orders, eliminating unphysical odd-order terms, and thus suppresses the negative sign problem coming from these interactions. 
%This reduces the computational cost considerably.
In particular, in the single-site DMFT, we find that the negative signs are absent.
%the proposed method completely resolves the sign problem, 
%as far as the two orbitals do not hybridize with each other.

%First, we show that without $\delta_2$, the weight for a configuration with odd number of non-density-type vertices (spin-flip and pair-hopping ones) vanishes in the case where $[\tilde{\mathcal{G}}_{0\sigma} ]_{12} = 0$.
%Since only $[\tilde{\mathcal{G}}_{0\sigma} ]_{11}$ and $[\tilde{\mathcal{G}}_{0\sigma} ]_{22}$ are non-zero, we need the even number of $\hat{c}^{\dagger}_{1\sigma}$, $\hat{c}^{\dagger}_{2\sigma}$, $\hat{c}_{1\sigma}$, and $\hat{c}^{\dagger}_{2\sigma}$ to have a non-zero thermal average of the products of them $\langle V[\hat{c}^{\dagger}_{1\sigma}, \hat{c}^{\dagger}_{2\sigma}, \hat{c}_{1\sigma}, \hat{c}^{\dagger}_{2\sigma}] \rangle_0$. 
%If the number of non-density vertices is odd, at least two of $\hat{c}^{\dagger}_{1\sigma}$, $\hat{c}^{\dagger}_{2\sigma}$, $\hat{c}_{1\sigma}$, and $\hat{c}^{\dagger}_{2\sigma}$ become a odd number
%This situation always satisfy when only the density-density terms exist in the configuration.
%However, if non-density-type vertices exist, they must be even number to satisfy the even number of $\hat{c}^{\dagger}_{1\sigma}$, $\hat{c}^{\dagger}_{2\sigma}$, $\hat{c}_{1\sigma}$, and $\hat{c}^{\dagger}_{2\sigma}$, otherwise the weight vanishes.
%When $\delta_2$ is non-zero, a configuration with odd number of non-density-type vertices has finite weight and therefore the ergodicity is satisfied with single-vertex insertion and single-vertex removal. 

In order to clue in our idea, let us look into Eq.~(\ref{Eq.Sint}) again. Suppose that there is no hybridization between the two orbitals, that is, $[\tilde{\mathcal{G}}_{0\sigma} ]_{12}=[\tilde{\mathcal{G}}_{0\sigma} ]_{21} = 0$.
Then we can easily see that, without $\delta_2$, the thermal average of the products of the Grassmann variables, $\langle V[\hat{c}^{\dagger}_{1\sigma}, \hat{c}^{\dagger}_{2\sigma}, \hat{c}_{1\sigma}, \hat{c}_{2\sigma}] \rangle_0$, can be finite only when 
the equality (number of $\hat{c}^{\dagger}_{i\sigma}$ in $V$) = (number of $\hat{c}_{i\sigma}$ in $V$) holds for each $i=1,2$ and $\sigma = \uparrow, \downarrow$.
%$V[\hat{c}^{\dagger}_{1\sigma}, \hat{c}^{\dagger}_{2\sigma}, \hat{c}_{1\sigma}, \hat{c}_{2\sigma}]$ contains an even number of the Gassmann variables for each spin and orbital component.
This condition is always satisfied when only the density-type vertices come in.
However, a single non-density-type vertex (spin-flip or pair-hopping) does not meet this condition, and therefore it must always appear in pair with another corresponding non-density-type vertex in order to have a finite contribution.
Nevertheless, when $\delta_2$ is non-zero, a configuration with the odd number of the non-density-type vertices can have a finite weight because of the constant $\alpha_{2\sigma}$.
While the presence of these odd-order terms is artificial, they are necessary to keep the ergodicity within the single-vertex update processes since in this case the number of the non-density-type vertices cannot be changed without passing through the odd-order terms.

%Given this situation, as for the update of non-density-type vertices, we only need to preform double-vertex insertion or double-vertex removal and we do not need to introduce $\alpha_2$ parameter in the conventional update scheme.
The above consideration motivates us to introduce double-vertex insertion or removal processes for the spin-flip and pair-hopping terms, where we insert or remove two non-density-type vertices at different imaginary times simultaneously.
With the double-vertex update processes we can sample over only the even-order terms with respect to the non-density-density interactions so that we can avoid the negative signs coming from the artificial odd-order terms.
The idea can apply to both the conventional and submatrix update algorithms.
While the double-vertex update dispenses with the additional parameter $\delta_2$ in the conventional fast update scheme, 
%and we can show 
%mathematically that there are no negative weights. The detail of the proof is elaborated in Appndix~\ref{sec_ap_proof}.
 in order to apply the submatrix update, we introduce another type of parameters,
%In the case of submatix update, we introduce an additional parameter 
\begin{eqnarray}
\alpha_{3\uparrow}(s) &=&  +s \delta_3 \nn
\alpha_{3\downarrow}(s) &=&  - s \delta_3
\end{eqnarray}
and 
\begin{eqnarray}
\alpha_{4\uparrow}(s) &=&  +s \delta_3 \nn
\alpha_{4\downarrow}(s) &=&  + s \delta_3
\end{eqnarray}
with $s=\pm1$ and a positive real number $\delta_3$. 
These parameters are needed to avoid the divergence of $\gamma$ function in Eq.~(\ref{Eq.gamfunc}).
We rewrite the action for the spin-flip and pair-hopping part as 
\begin{widetext}
\begin{eqnarray}
 S_{\rm{int}}^{\rm{non\mathchar`-dens.}}    =  \int^{\beta}_{0} d \tau  \sum_{s=\pm1} &\Biggl[&
    \underbrace{ \sum_{l=3,4}  \frac{J_{\rm H}}{4}
\bigl[ \hat{c}^{\dagger}_{1\uparrow}(\tau) \hat{c}^{\ }_{2\uparrow} (\tau)- \alpha_{l\uparrow}(s) \bigr ] 
\bigl[ \hat{c}^{\dagger}_{2\downarrow}(\tau) \hat{c}^{\ }_{1\downarrow}(\tau)  - \alpha_{l\downarrow}(s) \bigr ] }_{\text{\Large$\kappa=7$}} 
\nn
 &+&  \underbrace{ \sum_{l=3,4}  \frac{J_{\rm H}}{4}
\bigl[ \hat{c}^{\dagger}_{2\uparrow}(\tau) \hat{c}^{\ }_{1\uparrow} (\tau)- \alpha_{l\uparrow}(s) \bigr ] 
\bigl[ \hat{c}^{\dagger}_{1\downarrow}(\tau) \hat{c}^{\ }_{2\downarrow}(\tau)  - \alpha_{l\downarrow}(s) \bigr ] }_{\text{\Large$\kappa=8$}} 
\nn 
&+& \underbrace{ 
%\sum_{i \neq j} 
\sum_{l=3,4}  \frac{J_{\rm H}}{4}
\bigl[ \hat{c}^{\dagger}_{1\uparrow}(\tau) \hat{c}^{\ }_{2\uparrow} (\tau)- \alpha_{l\uparrow}(s) \bigr ] 
\bigl[ \hat{c}^{\dagger}_{1\downarrow}(\tau) \hat{c}^{\ }_{2\downarrow}(\tau)  - \alpha_{l\downarrow}(s) \bigr ] }_{\text{\Large$\kappa=9$}}
\nn 
&+&
\underbrace{ 
%\sum_{i \neq j} 
\sum_{l=3,4}  \frac{J_{\rm H}}{4}
\bigl[ \hat{c}^{\dagger}_{2\uparrow}(\tau) \hat{c}^{\ }_{1\uparrow} (\tau)- \alpha_{l\uparrow}(s) \bigr ] 
\bigl[ \hat{c}^{\dagger}_{2\downarrow}(\tau) \hat{c}^{\ }_{1\downarrow}(\tau)  - \alpha_{l\downarrow}(s) \bigr ] }_{\text{\Large$\kappa=10$}}
\Biggr].
\label{Eq.Sint_ndd}
\end{eqnarray}
\end{widetext}
%$S_0$ and the density-density part of $S_{\rm int}$ does not change. 
%Note that the sum over $s$ and $l$ returns to the original action without the constant term. 
%Therefore, the weight for a configuration with odd number of non-density-type vertices still vanishes even if we introduce $\alpha_3$ and $\alpha_4$ parameters.
%where $\kappa = 1\mathchar`-6$ denote density-density terms.
The idea behind this form of the additional parameters $\alpha_{3\sigma}$ and $\alpha_{4\sigma}$ is to eliminate the weight of the odd-order terms, as one can easily verify it by seeing that the sum over $s$ and $l$ for the each term on the right hand side of Eq.~(\ref{Eq.Sint_ndd})  reproduces the original action for the spin-flip and pair-hopping terms without the additional constant.
We use the same $S_0$ and the density-density part of $S_{\rm int}$ as those in Eqs.~(\ref{Eq.S0_multi}) and (\ref{Eq.Sint}), 
where we assign $\kappa=1\mathchar`-6$ to the density-density interactions in $S_{\rm int}$.
%We also note that to realize the even number of $\hat{c}^{\dagger}_{1\sigma}$, $\hat{c}^{\dagger}_{2\sigma}$, $\hat{c}_{1\sigma}$, and $\hat{c}^{\dagger}_{2\sigma}$,  
In the update,
the $\kappa=7$ vertex has to be 
%inserted as a pair
paired
 with $\kappa=8$ vertex. In the same way, the $\kappa=9$ vertex has to be 
%inserted as a pair 
paired
with $\kappa=10$ vertex.
In principle, $\delta_3$ 
%can be 
is
arbitrary 
as far as it is nonzero, however, 
%considering the fact that $\delta_3$ is introduced to incorporate the double-vertex update in submatrix update and that we do not need it for the conventional update, we set $\delta_3$ to be very small $\sim 10^{-4}$.
a small value is preferable because in the $\delta_3 \rightarrow 0$ limit, we can prove mathematically that the negative signs are absent (see Appendix~\ref{sec_ap_proof}). 
We set $\delta_3$ to be $\sim 10^{-4}$, and with this small value, we do not encounter the negative signs as will be shown in Sec.~\ref{sec_result_db}.
A large value of $\delta_3$ will increase the matrix size and produce the negative signs.

%In this case, $\kappa$ runs from 1 to 14. 
%$\kappa =1,2$ denote $U$ terms, $\kappa =3,4$ denote $U'$ terms, $\kappa =5,6$  denote $U'-J_{\rm H}$ terms, 
%$\kappa =7,8, 9,$ and 10  denote spin-flip terms, 
%$\kappa =11,12, 13$  and 14 denote pair-hopping terms, respectively.
% with $\alpha_3$ parameters, 
%$\kappa =13,14$  denote pair-hopping terms with $\alpha_4$ parameters, respectively.
$f$ function for the non-density-type vertices is modified 
%as
to
\begin{eqnarray}
 f_{\kappa\sigma}(l,s) = \left \{
 \begin{array}{ll} 
    \frac{\alpha_{l\sigma} (s)} {\alpha_{l\sigma} (s)- 1 }   &  s =\pm1 \\
    1  & s = 0 
 \end{array}
 \right . , 
\end{eqnarray}
%for $\kappa = 7,9,11,$ and 13,  
%and 
%\begin{eqnarray}
% f_{\kappa\sigma}(s) = \left \{
% \begin{array}{ll} 
%    \frac{\alpha_{4\sigma} (s)} {\alpha_{4\sigma} (s)- 1 }   &  s =\pm1 \\
%    1  & s = 0 
% \end{array}
 %\right . , 
%\end{eqnarray}
for $\kappa = 7$-$10$ and $l=3,4$.
% and correspondingly, $V_{\kappa} = J_{\rm H}/2$.
Correspondingly we define $V_{\kappa} = J_{\rm H}/2$, with which
%Then, 
the partition function is given in the same form as Eq.~(\ref{Eq.Z_multi}).

%As for
At the insertion update, we 
%set the ratio $R$ for the double-vertex update, and single-vertex update is done with the ratio $1-R$. In the case of the double-vertex update,
propose the double vertex with a probability $R$, and the single vertex with $1-R$. When the double-vertex update is selected,
 we randomly choose 
 %the vertex type from $\kappa=7$ to $\kappa=14$ for the first vertex.
either pair of (7,8) or (9,10).
%, (8,9), (8,10), (11,13), (11,14), (12,13), (12,14) \}.}
%If the type of the chosen vertex is 7 or 8 for example, we choose randomly the second vertex from $\kappa=9,10$. 
Then, we pick two imaginary times from the range $[0,\beta)$ and 
%choose the
assign
the $l$ value ($l=3,4$) 
and
 auxiliary spin orientations ($s=\pm1$) for 
%these pairs.
each vertex in the pair.
%Finally
Eventually
 a proposal ratio for inserting a certain pair of the non-density-type vertices is $R/32 \times (d \tau /\beta )^2$.
As for the removal update, we first pick randomly one of the existing vertices.
If the chosen vertex is of density-density type, we propose the single-vertex removal.
Otherwise, we propose the double-vertex removal: If the type of the chosen vertex is 7, for example, 
%then 
we additionally choose one 
%$\kappa=9,10$ 
vertex from the existing $\kappa=8$ vertices  with a probability $1/m_{\kappa=8}$ with $m_{\kappa=8}$
% = m_{\kappa=7,8}$
 being the number of $\kappa=8$ vertices in the configuration.
 Then a proposal ratio for removing a (7,8) pair is $\frac{2}{n m_{\kappa=7}}$, where $n$ is the number of existing vertices of all kinds.
 The factor of 2 in the numerator comes from the sum of probability for the case where the first-chosen vertex is of $\kappa=7$ and $\kappa=8$.
Note that $m_{\kappa=7} = m_{\kappa=8}$ and $m_{\kappa=9} = m_{\kappa=10}$ always hold during the simulation. 
With $X=\frac{16K_{\kappa=7}^2}{R(n+2)(m_{\kappa=7}+1)}$, 
the acceptance ratio 
%for the insertion and removal for
concerning
 $(7,9)$-pair vertices is 
%given as 
\begin{eqnarray}
P(C_n \! \rightarrow \! C_{n+2}) = {\rm min} \left( X
\prod_{\sigma} \frac{ {\rm det} A_{\sigma}(C_{n+2})}{{\rm det} A_{\sigma}(C_n)}, 1   \right)
\label{Eq.add}
\end{eqnarray}
for the addition 
%of the vertices,
process
 and 
\begin{eqnarray}
P(C_{n+2} \! \rightarrow \! C_{n}) = {\rm min} \!  \left( \frac{1}{X}
\prod_{\sigma} \frac{{\rm det} A_{\sigma}(C_n)}{ {\rm det} A_{\sigma}(C_{n+2})} , 1   \right)
\label{Eq.rem}
\end{eqnarray} 
 for the removal 
%of the vertices.
process.
The acceptance ratios for the insertion and the removal of the other vertex pairs are calculated in the same way.
%If the double-vertex update corresponds to $(s^{k} =0 \rightarrow s^{k+1}=\pm1, s'^{k} =0 \rightarrow s'^{k+1}=\pm1)$ or $(s^{k} = s^{0}=\pm1 \rightarrow s^{k+1}=0, s'^{0}= s'^{0}=\pm1 \rightarrow s'^{k+1}=0)$ [ $(s^{k} =\pm1 \rightarrow s^{k+1} =0 = s^{0}, s'^{0}=\pm1 \rightarrow s'^{k+1}=0= s'^{0})$ ], we enlarge [shrink] the $\Gamma$ matrix by two rows and two columns. 
%Otherwise, the update corresponds to  $(s^{k} =s^{0} = \pm1 \rightarrow s^{k+1} =0 , s'^{0}=\pm1 \rightarrow s'^{k+1}=0= s'^{0})$ or $(s^{k} = \pm1 \rightarrow s^{k+1} =0 =s^{0}  , s'^{0}= s'^{0}=\pm1 \rightarrow s'^{k+1}=0)$, and we shrink the $\Gamma$ matrix by one rows and one columns for the spin that comes back to the initial spin configuration and simultaneously enlarge the $\Gamma$ matrix by one rows and one columns for the other spin.

Suppose a pair of the auxiliary spins, $(s_p^k, s_q^k)$, is proposed to change to $({s'}_p^k, {s'}_q^k)$ by the double-vertex update.
As far as $p$-th and $q$-th spins have not been changed in the previous $(k-1)$ steps, the change of the type $(0,0) \rightarrow (\pm1,\pm1)$ (insertion) or $(\pm1,\pm1) \rightarrow (0,0)$ (removal) will enlarge the $\Gamma$ matrix by two rows and two columns if accepted.
If both the $p$-th and $q$-th spins have already been flipped from 0 to $\pm1$ (insertion), the change at the $k$-th step is of the type $(\pm1,\pm1) \rightarrow (0,0)$ (removal) and the $\Gamma$ matrix will shrink by two rows and two columns if accepted, 
since both the $p$-th and $q$-th spins return to the original orientations ($s=0$).
Otherwise, one of the two spins, say the $p$-th spin, has been changed in the previous $(k-1)$ steps while the other (the $q$-th spin) has not.
In this case, the change is of the type $(\pm1,\pm1) \rightarrow (0,0)$ (removal) and in the $\Gamma$ matrix one row and one column will be  added for the $q$-th spin while one row and one column concerning the $p$-th spin will be removed if accepted.

Finally, we comment on the three-orbital case.
Suppose that there is no hybridization among the orbitals.
In this case, on top of the double-vertex update,  we will need the triple-vertex update, 
where, three spin-flip interactions involving the orbital pairs (1,2), (2,3), and (3,1), for example,  are inserted or removed.

\subsection{Multi-orbital and multi-site case}\label{sec_mo_ms}
It is straightforward to extend the above-described algorithm, 
both the single-vertex and double-vertex updates,
 to the multi-orbital and multi-site impurity problem. We only need to define a ``{\it generalized orbital}'' which specifies the site and the orbital simultaneously.
With these ``{\it generalized orbitals}'', we can employ the same method described in Sec.~\ref{subsec_multi}. 
For example, when we consider two-orbital and two-site case, the ``{\it generalized orbital}'' runs from 1 to 4:    
``{\it Generalized orbital}'' 1, 2, 3, and 4 denote the orbital 1 at the site 1, the orbital 2 at the site 1, the orbital 1 at the site 2, and 
the orbital 2 at the site 2, respectively. 
The Weiss function becomes a matrix with respect to the ``{\it generalized orbitals}'' and includes the off-site processes, e.g., $\bigl[ \tilde{\mathcal{G}} ^{-1}_{0\sigma} (\tau - \tau') \bigr]_{13}$. 
It also should be noted that, for the 
multi-orbital Hubbard model,
 the interactions exist only within the ``{\it orbital}'' 1 and the ``{\it orbital}''  2, and within the ``{\it orbital}'' 3 and the ``{\it orbital}'' 4.

\section{results}\label{sec_result}
Here, we show numerical results for the 2D two-orbital Hubbard model. 
We consider two degenerate orbitals on a square lattice with only the nearest neighbor intra-orbital hopping $t$, which is used as 
%an energy unit ($t=1$).
the unit of energy, i.e., $t=1$.
The electron density is set to be half filling.
We implement the cellular DMFT with a four-site cluster,
in which the impurity problem has $2\times4=8$ degrees of freedom in total, and compare the results with those of the single-site DMFT to elucidate the effect of short-range spatial correlations.

The impurity problem is solved by the CT-INT method described in the previous section, 
where the Legendre orthogonal polynomials expansion of the imaginary-time Green's function is employed as a ``noise filter".~\cite{PhysRevB.84.075145}
We restrict ourselves to the paramagnetic and para-orbital solution to clarify the nature of the Mott metal-insulator transition.
We explicitly treat the spin-flip and pair-hopping terms (the SU(2)-symmetric Hamiltonian) and compare the result with that of the $Z_2$-symmetric Hamiltonian. 
%\tr{Using 512 parallelization, it takes about one hour to preform one self-consistent loop for SU(2)-symmetric Hamiltonian at $T/t=0.05$ and $U/t = 5.4$, where the calculation becomes most expensive in the present study.}

For the SU(2)-symmetric Hamiltonian at $T/t = 0.05$ and $U/t = 5.4$, where the calculation is severest in the present study, the average expansion order of the interaction vertices reaches $\sim$ 740 and  
we take 1,536,000 QMC steps to solve the impurity problem. In this case, it takes about one hour with 512-core parallelization (clock frequency: 2.90GHz) to perform one self-consistent loop.

\subsection{Comparison between single-vertex and double-vertex updates}\label{sec_result_db}
\begin{figure}[tbp]
\vspace{0cm}
\begin{center}
\includegraphics[width=0.4\textwidth]{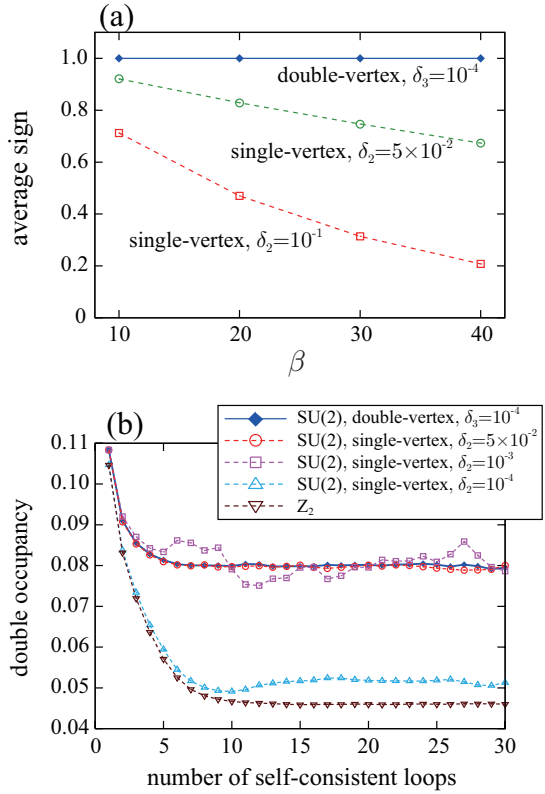}
\caption{
(Color online) 
(a) The 
%norm
average sign 
 for  the SU(2)-symmetric Hamiltonian obtained within the single-site DMFT. 
Filled (open) symbols show the results with the double-vertex (single-vertex) update for the spin-flip and pair-hopping terms. 
(b) The double occupancy for each orbital for the SU(2)-symmetric Hamiltonian at $\beta=20$ as a function of the number of the self-consistent loops, where we employ the single-site DMFT.
For comparison, we also show the results for the $Z_2$-symmetric Hamiltonian.  We start the self-consistent loop from the non-interacting limit, 
%i.e., no self-energy and no vertices, 
and we fully update the Weiss function at each loop.
In the QMC simulation, 320,000 measurements are done. 
The calculation is performed with $U = 6t$, and $U' = 3t$ and $J_{\rm H}=1.5 t$ both for the panels (a) and (b).
} 
\label{fig_nega}
\end{center}
\end{figure}

Before 
%we show 
going to
the physical results for the 2D two-orbital Hubbard model, we 
%show
demonstrate
 how much the negative 
%sign is
signs are
 reduced by employing the double-vertex update for the spin-flip and pair-hopping terms.
The calculation is performed 
%with 
at
$U = 6t$, 
%and 
$U' = 3t$ and $J_{\rm H}=1.5 t$.
Fig.~\ref{fig_nega}(a) shows the single-site DMFT results 
%for the norm 
of the average sign
for the SU(2)-symmetric Hamiltonian at several temperatures.
As can be seen, the double-vertex update 
%overcomes the sign problem,  while ratio of the negative sign increases as the temperature goes down or as the $\delta_2$ value increase in the case of the single-vertex update.
%One might think from Fig.~\ref{fig_nega}(a) that if we further decrease $\delta_2$, we can get rid of the sign problem. 
always gives the average sign of 1, eliminating the negative signs completely. On the other hand, the single-vertex update suffers from the negative signs, which become severer as the temperature decreases. Since the slope in Fig.~\ref{fig_nega}(a) is more modest for the smaller value of $\delta_2$, one might think
that if we further decrease $\delta_2$, we can get rid of the sign problem. 
%It is true that smaller $\delta_2$ leads to smaller ratio of negative sign, 
However, if $\delta_2$ is too small, the calculation becomes unstable, 
% (Fig.~\ref{fig_nega}(b)):
as seen in Fig.~\ref{fig_nega}(b):
The result with $\delta_2 = 10^{-3}$ strongly fluctuates around the right value (red and blue curves)
%for the double occupancy 
$\sim 0.08$, and 
%what is worse, we do not reach the right solution when $\delta_2 = 10^{-4}$, rather the results with $\delta_2 = 10^{-4}$ are close to the results for $Z_2$-symmetric Hamiltonian.
%The reason for the unstable results is ascribed to the fact that the weight for a configuration with odd number of non-density-type vertices becomes smaller and smaller as $\delta_2$ decreases and that it becomes harder and harder to change the order of non-density-type vertices.
%Therefore, in order to get a good result with the single-vertex update, we can not avoid the sign problem. 
%On the other hand, the double-vertex update method is free from the sign problem and is much more efficient.
for $\delta_2 = 10^{-4}$ even the average value of the solution deviates from the right one. The result with $\delta_2 = 10^{-4}$ is rather close to the result with the $Z_2$-symmetric Hamiltonian.
This is reasonable because the reduction of $\delta_2$ suppresses the flip to the odd-order non-density-type terms: Since we start from the non-interacting limit (0th order), the smaller $\delta_2$ lessens the chance to have a finite-order non-density-type terms, resulting in a double-occupancy value similar to the $Z_2$-symmetric one. 
Therefore, if we want an accurate and stable result with the single-vertex update, we need to use a substantial value for $\delta_2$, which inevitably causes negative signs.
%a considerable amount of negative signs.
On the other hand, in the double-vertex update, 
the accuracy does not essentially depend on the choice of $\delta_3$, and as far as we use a small value for $\delta_3$, we see that the average sign is always one.
The computational time highly depends on the average sign: If the average sign is 0.5, we need a twice larger calculation to get the same effective sampling numbers as that of (average sign) = 1 case. Therefore, the double-vertex update saves the computational time significantly.

\begin{figure}[tbp]
\vspace{0cm}
\begin{center}
\includegraphics[width=0.45\textwidth]{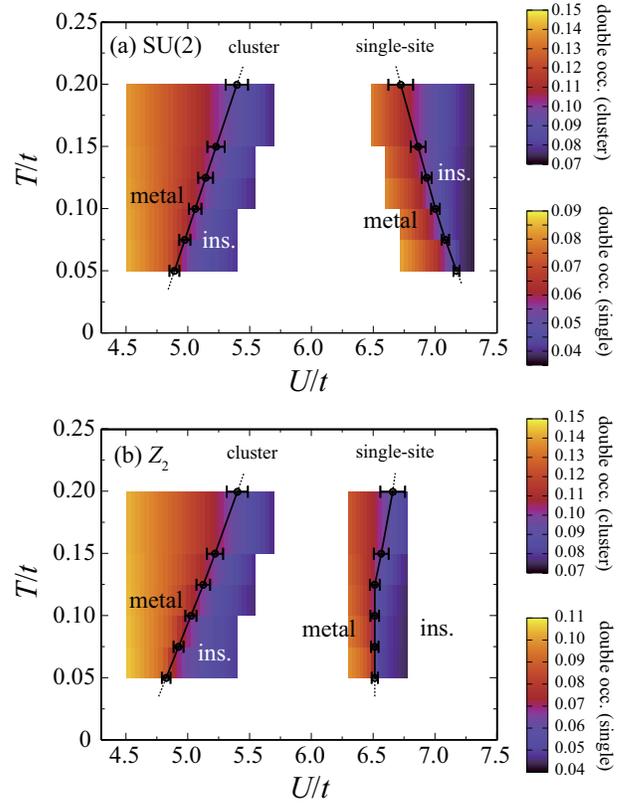}
\caption{(Color online) Phase diagrams obtained by the cDMFT and single-site DMFT for (a) the SU(2)-symmetric Hamiltonian and (b) the $Z_2$-symmetric Hamiltonian with $J_{\rm H}/ U = 1/6$ and $U' = U- 2J_{\rm H}$.
Color contour plots show the double occupancy for each orbital,  
where the data between the calculated points are estimated by a linear interpolation.
%\tr{The solid lines shows the phase boundary starting from the metallic solution or the crossover line. Within the present  resolution, we could not identify the critical end point precisely, hence, we do not distinguish the phase boundary from the crossover line.}
The solid lines show the phase boundary at which the metallic solution becomes unstable, or the crossover line determined by the maximal point of the first derivative of the double occupancy as a function of $U$. Within the present resolution, we could not determine the critical end point precisely.
%\tg{
%For the temperature region used in the calculation, the region where the metallic and insulating solution coexists are 
%very small ($\sim 0.1 t$), if exists. 
%}
} 
\label{fig_pd}
\end{center}
\end{figure} 

\begin{figure}[tbp]
\vspace{0cm}
\begin{center}
\includegraphics[width=0.35\textwidth]{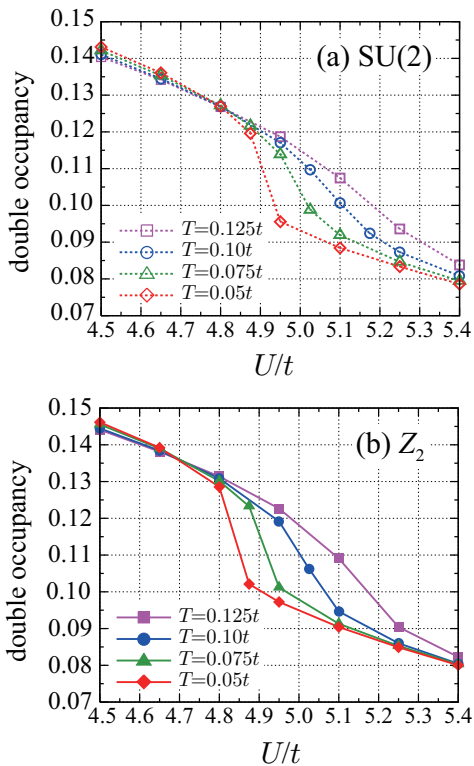}
\caption{(Color online) The cDMFT results for the double occupancy 
for each orbital for (a) the SU(2)-symmetric Hamiltonian and (b) the $Z_2$-symmetric Hamiltonian with $J_{\rm H}/ U = 1/6$ and $U' = U- 2J_{\rm H}$.
%The comparison between them is shown in the panel (c), where the same symbols as in the panels (a) and (b) are used.
The lines are guides to the eye. The sizes of the error bars are slightly small compared to the sizes of symbols. 
} 
\label{fig_db}
\end{center}
\end{figure} 

\begin{figure*}[tbp]
\vspace{0cm}
\begin{center}
\includegraphics[width=0.92\textwidth]{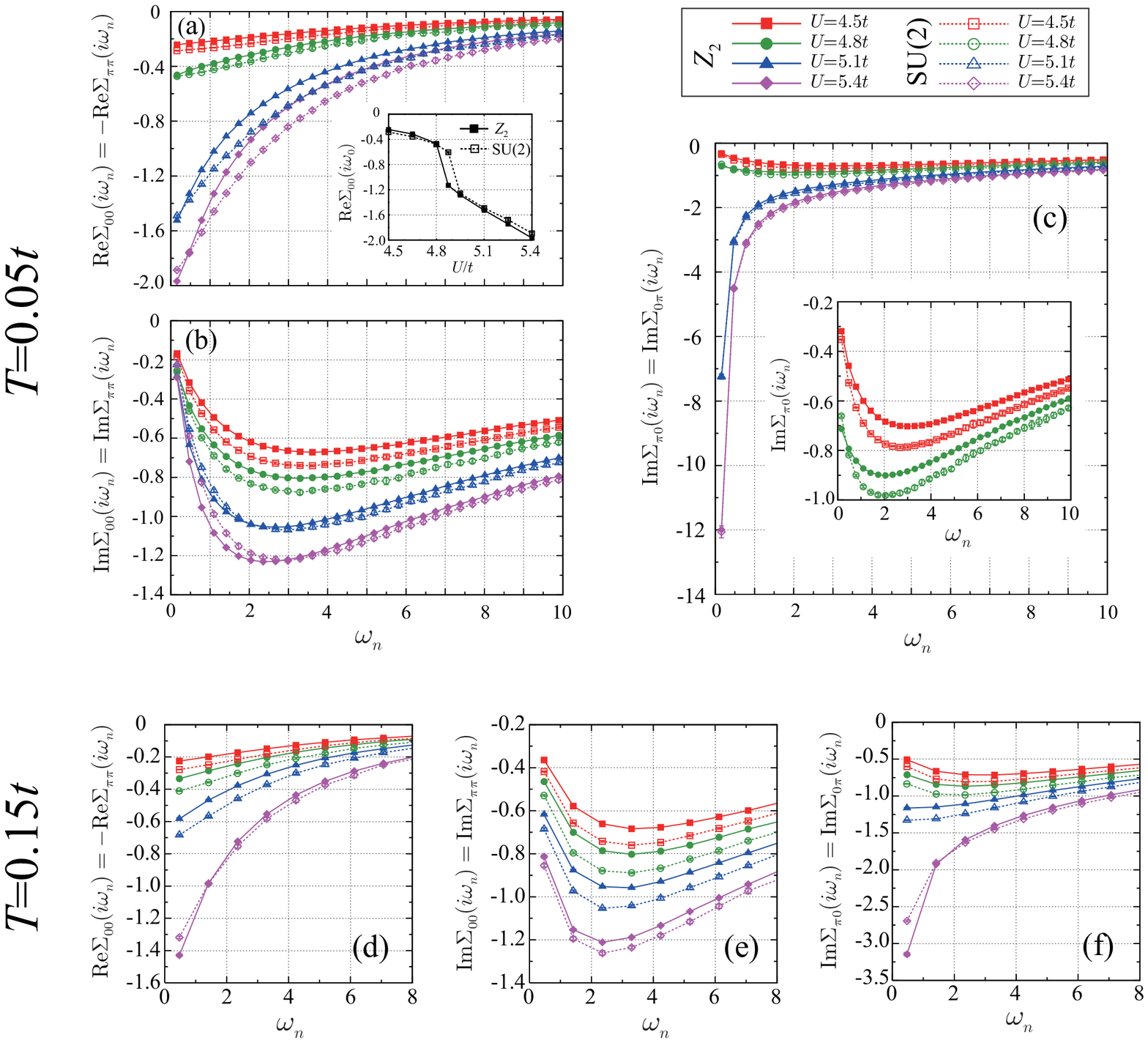}
\caption{(Color online) 
(a)-(c) The cDMFT results for the self-energies at $T = 0.05t$.
%The cDMFT results for the self-energies 
%at $T = 0.05 t$ [panels (a)-(c)] and at $T = 0.15 t$ [panels (d)-(f)].
%Panels (a), (d), and (g) [(b), (e), and (h)] show the results for SU(2)- [$Z_2$-]symmetric Hamiltonian. The comparison between them is shown in the panels (c), (f), and (i). 
The panels (a), (b), and (c) show the data for the real part of the self-energy at the $(0,0)$ momentum, ${\rm Re} \Sigma_{00} (i\omega_n) = -{\rm Re} \Sigma_{\pi\pi} (i\omega_n)$, 
its imaginary part, ${\rm Im} \Sigma_{00} (i\omega_n) = {\rm Im} \Sigma_{\pi\pi} (i\omega_n)$, and the imaginary part of the self-energy at the $(\pi,0)$ momentum, ${\rm Im} \Sigma_{\pi0} (i\omega_n) = {\rm Im} \Sigma_{0\pi} (i\omega_n)$, respectively.
Due to the particle-hole symmetry, ${\rm Re} \Sigma_{\pi0} (i\omega_n) = {\rm Re} \Sigma_{0\pi} (i\omega_n) = 0 $.
The inset of the panel (a) show the real part of the self-energy for the $(0,0)$ momentum at the first Matsubara frequency, ${\rm Re} \Sigma_{00} (i\omega_0) = -{\rm Re} \Sigma_{\pi\pi} (i\omega_0)$, as a function of the Hubbard interaction $U$. 
The data for   ${\rm Im} \Sigma_{\pi0} (i\omega_n) = {\rm Im} \Sigma_{0\pi} (i\omega_n)$ at $U/t=4.5, 4.8$ are zoomed in the inset of the panel (c). 
(d)-(f) The same as (a)-(c) but at $T=0.15t$.
%The definitions of the symbols are used in common in all the panels.  
The sizes of the error bars for the data for the $Z_2$-symmetric Hamiltonian are within those of the symbols. The lines are guides to the eye. 
} 
\label{fig_SE}
\end{center}
\end{figure*}

\subsection{Phase diagram}
Figures~\ref{fig_pd}(a) and \ref{fig_pd}(b) show the phase diagrams with respect to the temperature $T$ and the interaction $U$ for the SU(2)-symmetric Hamiltonian and the $Z_2$-symmetric one, respectively, where the ratio between Hund's coupling $J_{\rm H}$ and the Hubbard interaction $U$ is set to be $J_{\rm H}/ U = 1/6$, and $U' = U- 2J_{\rm H}$. 
The ratio $J_{\rm H}/ U = 1/6$ is close to that of the transition metal oxides,~\cite{PhysRevB.86.085117,PhysRevB.86.165105} typical multi-orbital strongly correlated materials.
%Hereafter, the results for $J_{\rm H}/ U = 1/6$ are given.
The color contour plot indicates the double occupancy obtained by the solution approached from the metallic side.
The raw data of the double occupancy are shown in Fig.~\ref{fig_db}.
The transition from a metallic state to the Mott insulating state can be identified by the abrupt change in the double occupancy. 
As the temperature increases, the change gets smoother and goes on to a crossover-like behavior, where we determine the crossover line by the maximal point of the first derivative of the double occupancy curves as a function of $U$.
In Figs.~\ref{fig_pd}(a) and \ref{fig_pd}(b), we show thus-estimated phase boundary $U_{c_2}$ or the crossover line of the Mott metal-insulator transition obtained by the single-site and cellular DMFTs.

First, we comment on the single-site DMFT results. 
%As for 
In the SU(2)-symmetric case, the critical interaction strength increases as the temperature decreases, which reflects the fact that the paramagnetic insulating state has a larger entropy than the metallic state, as in the single-orbital case. 
On the other hand, $U_{c_2}$ for $Z_2$-symmetric Hamiltonian is almost unchanged with respect to the temperature while in the crossover region ($T \gtrsim 0.12t$), the crossover line shifts to a larger $U$ as the temperature increases.
%One of the reasons for the different slope is the difference in the ground-state degeneracy in the atomic limit differs between SU(2)- and $Z_2$-symmetric cases. 
%In the latter case, the ground states, where each orbital has one electron with a spin oriented to the same direction ($S=1$, $S_z \pm1$),  are doubly degenerate, while in the former case they are triply degenerate ($S=1$, $S_z = 1,0,-1$).
The different slopes between SU(2) and $Z_2$ come from their different ground-state degeneracy in the atomic limit where each orbital has one electron with a spin oriented to the same direction ($S=1$).
In the SU(2) case the ground state is triply degenerate ($S_z = 1,0,-1$) while in the $Z_2$ case it is doubly degenerate ($S_z=\pm 1$).
Hence, the insulating state in the SU(2)-symmetric Hamiltonian has a larger entropy than that in the $Z_2$-symmetric Hamiltonian, 
%leading to a
accounting for the tendency to have a negative slope of the phase boundary in the 
%former
SU(2) case.
Furthermore, in the metallic region for the $Z_2$-symmetric Hamiltonian, since the system is locked into the states with $S_z =\pm1$ 
due to a strong Hund's coupling, the Kondo screening is 
%absent
inefficient,~\cite{EPJB.44.217} while it 
%is present
works
 in the SU(2)-symmetric Hamiltonian 
%and 
as well as in the single-orbital 
%case.
one.
Therefore, the metallic state in the 
%former
multi-orbital $Z_2$
 case has a larger entropy 
%compared to
than
 that in the 
%latter
multi-orbital SU(2) and single-orbital
 cases. 
%This might explain the different \tr{slopes} between single-orbital and two-orbital 
%\tr{$Z_2$}
%cases, although, in both cases, the ground states 
%%at
%\tr{in}
% the atomic limit are doubly degenerate, which would give a similar entropy in the insulating region. 
Since in the atomic limit both the single-orbital and multi-orbital $Z_2$ Hamiltonians have the same ground-state degeneracy of two, which would give a similar entropy in the insulating region, the above-mentioned difference in the metallic state would explain the positive slope in the $Z_2$ case.
Notice also that the $Z_2$-symmetric Hamiltonian significantly overestimates the tendency toward the insulator
compared to the SU(2)-symmetric one.

We now turn to the cellular DMFT results. 
Due to the short-range spatial correlations, the critical interaction strength for the Mott transition considerably decreases.  
It is interesting to note that the difference in the critical interaction strength between SU(2)- and $Z_2$-symmetric Hamiltonians is much smaller than that in the single-site DMFT.
By comparing Figs.~\ref{fig_db}(a) and \ref{fig_db}(b), we find that the $Z_2$-symmetric Hamiltonian
%According to Fig.~\ref{fig_db}(c), where the double occupancy is compared between SU(2) and $Z_2$, the latter 
overestimates the tendency toward the insulator while the difference of the critical interaction is less than $0.1t$.
In contrast to the single-site DMFT results, the slopes of the phase boundary 
in Fig.~\ref{fig_pd}
are also similar between the SU(2)- and $Z_2$-symmetric Hamiltonians: The critical interaction strength decreases as the temperature decreases in both cases. 
In the SU(2)-symmetric Hamiltonian, in analogy with the single-orbital case,~\cite{PhysRevLett.101.186403} this would be attributed to the entropy reduction of the insulating phase by
the formation of the inter-site singlets within the cluster. 
%in the SU(2)-symmetric Hamiltonian.
In the $Z_2$-symmetric case, the Ising-type antiferromagnetic spin alignment would be favored in the cluster and thus 
the insulating phase has a smaller entropy than that in the single-site DMFT.
%the entropy per site decreases compared to the single-site case.
To confirm these scenarios, it would be interesting to see the inter-site 
spin-spin correlation functions, which is however beyond the scope of the present study.
%To confirm these scenarios, we need to calculate inter-site spin-spin correlation function, which is a future problem.}

\subsection{Self-energy }
To investigate the nature of the transition, we plot in Fig.~\ref{fig_SE}(a)-\ref{fig_SE}(i) the raw data of the intra-orbital self-energy against the Matsubara frequency $\omega_n=(2n+1)\pi T$ for $U/t = 4.5$, $4.8$, $5.1$, and $5.4$ at the temperature $T = 0.05 t$.
%As a complementary information, 
%In addition,
%we 
%give the data for
%show
%the self-energy at a higher temperature $T=0.15t$ in Appendix~\ref{sec_ap2}.
The self-energy is diagonal with respect to the orbital and two orbitals give the same self-energy, while it has a momentum dependence. 
Figures~\ref{fig_SE}(a), 4(b), and 4(c) show the real part of the self-energy at the $(0,0)$ momentum ${\rm Re} \Sigma_{00} (i\omega_n) = -{\rm Re} \Sigma_{\pi\pi} (i\omega_n)$, its imaginary part ${\rm Im} \Sigma_{00} (i\omega_n) = {\rm Im} \Sigma_{\pi\pi} (i\omega_n)$, and the imaginary part of the $(\pi,0)$ component ${\rm Im} \Sigma_{\pi0} (i\omega_n) = {\rm Im} \Sigma_{0\pi} (i\omega_n)$, respectively. 
Note that the real part of the $(\pi,0)$ and $(0,\pi)$ components vanish due to the particle-hole symmetry.

First, we remark several 
%common features between 
features common to both
SU(2) and $Z_2$ results.
At the noninteracting limit $U/t = 0$, the Fermi surface exists at the $(\pi,0)$ momentum while the $(0,0)$- [$(\pi,\pi)$-]momentum state is occupied (unoccupied). 
In the Mott insulating state, this Fermi surface disappears at the $(\pi,0)$ momentum
%and ${\rm Im} \Sigma_{\pi0} (\omega \rightarrow 0)$ diverges,
due to the divergence of ${\rm Im} \Sigma_{\pi0} (\omega \rightarrow 0)$,
as can be seen from Fig.~\ref{fig_SE}(c). 
In the metallic region close to the Mott transition, the $(\pi,0)$-momentum self-energy does not go to zero but to a finite value as $\omega \rightarrow 0$, which is a sign of a bad metal.  
To investigate whether this bad metallic behavior is intrinsic or it becomes a good metal at lower temperatures requires a huge computational cost and is intractable at present.  
At the Mott transition, we see an abrupt change in ${\rm Re} \Sigma_{00} (i\omega_n)$ and ${\rm Re} \Sigma_{\pi\pi} (i\omega_n)$ 
(the inset of Fig.~\ref{fig_SE}(a)), 
which can also be used to determine the transition point. 
The similar change in ${\rm Re} \Sigma_{00} (i\omega_n)$ and ${\rm Re} \Sigma_{\pi\pi} (i\omega_n)$ is also seen in the cellular DMFT results for the 2D single-band Hubbard model on the square lattice.~\cite{PhysRevLett.101.186403}
On the other hand, through the Mott transition, we do not find any anomaly in the imaginary part of the self-energy at $(0,0)$ and $(\pi,\pi)$ momentum  [${\rm Im} \Sigma_{00} (i\omega_n)$ and ${\rm Im} \Sigma_{\pi\pi} (i\omega_n)$], where the Fermi surface does not exist even in the metallic state at small $U$.

%In Figs. \ref{fig_SE}(c), (f), and (i), we compare the self-energy between SU(2) and $Z_2$ cases at $U/t=4.5, 4.8, 5.1$ and $5.4$. 
We now turn to the comparison of the self-energy at $T = 0.05 t$ between SU(2) and $Z_2$ cases at $U/t=4.5, 4.8, 5.1$ and $5.4$.
For these values of interaction, the both types of Hamiltonian give a solution on the same side of the metal-insulator transition (see Fig.~\ref{fig_db} and the inset of Fig.~\ref{fig_SE}(a)),
and the difference in the resultant self-energies is at most $\sim20$\%.
%, which is also true at 
%the
%higher temperature $T=0.15t$ [see Figs.~\ref{fig_SE_high}(a), \ref{fig_SE_high}(b), and \ref{fig_SE_high}(c)].
A qualitative difference between SU(2) and $Z_2$ results can be seen only in the vicinity of the transition point: For example, for $U/t=4.875$ the SU(2)-symmetric Hamiltonian still remains 
%in
to give the metallic state while the $Z_2$-symmetric Hamiltonian incorrectly gives an insulating solution.

%\section{Self-energy at $T=0.15t$}\label{sec_ap2}
Finally, 
%to investigate whether the smallness of the difference in self-energy between $Z_2$- and SU(2)-symmetric Hamiltonian also 
%holds for higher temperature or not,  
we show the self-energy at $T=0.15t$ in 
Figs.~\ref{fig_SE}(d), \ref{fig_SE}(e), and \ref{fig_SE}(f), 
where the crossover behavior from the metal to the insulator is seen. 
 As is expected, the diverging 
%behaviors
behavior
 of  ${\rm Im} \Sigma_{\pi0} (i\omega_n)$ and ${\rm Im} \Sigma_{0\pi} (i\omega_n)$   
 is much more moderate compared to 
 that at $T=0.05 t$ 
 (Figs.~\ref{fig_SE}(a)-(c)). 
As for the difference between the results for the SU(2)-symmetric Hamiltonian and those for the $Z_2$-symmetric Hamiltonian, 
 generally 
the self-energies for the SU(2)-symmetric Hamiltonian are larger in magnitude, 
except for ${\rm Re} \Sigma_{00} (i\omega_0)$ and ${\rm Im} \Sigma_{\pi0} (i\omega_0)$. 
However, the difference is at most $\sim$ 20 \%.   
Similarly, we do not find any significant differences between the 
two types of Hamiltonian
for the other parameter sets which have been studied in this paper.
We however expect that these terms will give a 
substantial difference in 
two-particle quantities such as spin susceptibility (Ref.~\onlinecite{PhysRevB.74.155102}), which is left for future investigations.

\section{conclusion}\label{sec_sum}
We have incorporated the submatrix update into the CT-INT method and also developed the efficient sampling scheme, the double-vertex update, for the spin-flip and pair-hopping terms.
Using the developed method, we have performed the cellular DMFT study for the 2D two-orbital Hubbard model on the square lattice. 
We have shown that the short-range spatial correlations significantly reduce the critical interaction strength for the Mott transition. The transition is induced by the divergence of the imaginary part of
the $(\pi,0)$-momentum self-energy and simultaneously we see the abrupt change in ${\rm Re} \Sigma_{00} (i\omega_n) = -{\rm Re} \Sigma_{\pi\pi} (i\omega_n)$. 
While we see the overestimate of the tendency toward the insulator in the $Z_2$-symmetric Hamiltonian, the difference in the critical interaction value between with and without the spin-flip and pair-hopping terms are smaller for the cDMFT results than that in the single-site DMFT case in the parameter region we have studied.
When $J_{\rm H}$ is larger or a frustration is introduced,  the difference might be more significant even in the cDMFT, which is an open problem.

The present scheme has established a firm starting point for the multi-orbital cDMFT study. 
%Changing the filling from half-filling or increasing the number of orbital to three are feasible. It would be interesting to 
Calculations at away from half-filling and/or for more than two orbitals are feasible. It is also interesting
to study magnetism, superconductivity, orbital order, and so on, which we leave for future issues.

\begin{acknowledgements}
We would like to thank Philipp Werner, Giorgio Sangiovanni, and Nicolaus Parragh for fruitful discussions. 
This work was supported by Funding Program for World-Leading Innovative
R\&D on Science and Technology (FIRST program) on "Quantum Science
on Strong Correlation".
Y. N. is supported by the Grant-in-Aid from JSPS (Grant No. 12J08652).
The calculations were performed at the Supercomputer Center, ISSP, University of Tokyo. 
\end{acknowledgements}

\appendix
%\section{nature of the $A$ matrix }
\section{Calculation of the Green's function matrix $G$}
%Here, we note an important nature of the $A$ matrix which is utilized in the main text. The determinant of the $A$ matrix does not change even though we add an arbitrary number of vertices with ``{\it noninteracting}" spins ($s=0$), i.e., 
%\begin{eqnarray}
%{\rm det} A_{\sigma}(C_n) = {\rm det} \tilde{A}_{\sigma}(\tilde{C}_{n+k})
%\label{Eq.AAt}
%\end{eqnarray}
%where $k$ is an arbitrary number and 
%\begin{eqnarray}
% \tilde{C}_{n+k} = \{ (s_1 , \tau_1),  \cdots , (s_n, \tau_n) ,  (s_{n+1} =0 , \tau_{n+1})\cdots, (s_{n+k}=0, \tau_{n+k})\} 
%\ \  \tilde{C}_{n+k} = \{ \underbrace{ (s_1 , \tau_1),  \cdots , (s_n, \tau_n)}_{\text{\Large$C_n$}} ,  (0 , \tau_{n+1}),\cdots, (0, \tau_{n+k})\}.  \nonumber 
%\end{eqnarray}
%Furthermore, the $A$ matrix is related to \tr{the} Green's function matrix $G$ by $G_{\sigma} = A_{\sigma}^{-1} G_{0\sigma}$.
The Green's function matrix $G$ (or $\tilde{G}$) in Eq.~(\ref{Eq.Gam}) is related to the $A$ matrix by $G_{\sigma} = A_{\sigma}^{-1} G_{0\sigma}$. 
When the configuration $C'_n = \{ (s'_1 , \tau_1),  \cdots , (s'_n, \tau_n) \}$ differs from $C_n = \{ (s_1 , \tau_1),  \cdots , (s_n, \tau_n)\}$ in only the spin orientation, $A'_{\sigma} (C'_n)$ is related to $A_{\sigma}(C_n)$ via the Dyson equation
\begin{eqnarray}
 A_{\sigma}'^{-1} = A_{\sigma}^{-1} + (G_{\sigma}- I ) \Lambda_\sigma A_{\sigma}'^{-1}.  
\label{Eq.dyson}
\end{eqnarray}
Here $I$ is an $n \times n$ identity matrix and 
\begin{eqnarray}
\bigl[ \Lambda_{\sigma} \bigr]_{ij} = \delta_{ij} \frac{f_{\sigma}(s'_i) - f_{\sigma}(s_i)}{f_{\sigma}(s_i)}.
\end{eqnarray}
By setting $s'_i=0$ for all $i$ in Eq.~(\ref{Eq.dyson}), we obtain
 \begin{eqnarray}
 \left(  f_{\sigma} (s_j) - 1 \right )   [ G_{\sigma} ]_{ij} =    f_{\sigma} (s_j)  [ A_{\sigma}^{-1} ]_{ij}+ \delta_{ij} . 
\label{Eq.Ge}
\end{eqnarray}
If $s_j \neq 0$, we can use this efficient formula to calculate $ [ G_{\sigma} ]_{ij} $, 
otherwise, we need to compute $ [ G_{\sigma} ]_{ij} $ directly by 
\begin{eqnarray}
[ G_{\sigma} ]_{ij} = [ A_\sigma^{-1} ]_{ik} [G_{0\sigma}] _{kj}.  
\label{Eq.Gb}
\end{eqnarray}

\section{Absence of the sign problem within the double-vertex update}
\label{sec_ap_proof}
Here, we prove that the negative signs are absent within the
double-vertex update in the two-orbital systems, in a way similar to that employed in Ref.~\onlinecite{Werner_private_comm}
for the single-orbital Hubbard model.
We first consider the case of $\delta_3=0$ in Eq.~(\ref{Eq.Sint_ndd}).
Following Refs.~\onlinecite{Werner_private_comm} and \onlinecite{0305-4470-38-48-004},
 we introduce a chain representation for the non-interacting part of the impurity Hamiltonian,
\begin{eqnarray}
\tilde{\mathcal H}_0 = \sum_{i,\sigma}\sum_{r=0}^{\infty} \bigl[  \tilde{\epsilon}_{ir} \hat{d}^{\sigma\dagger}_{i,r} \hat{d}^{\sigma}_{i,r}  - 
t_{ir} ( \hat{d}^{\sigma\dagger}_{i,r+1} \hat{d}^{\sigma}_{i,r} + \hat{d}^{\sigma\dagger}_{i,r} \hat{d}^{\sigma}_{i,r+1} )   \bigr],  \nn 
\end{eqnarray}
where $\hat{d}^{\sigma\dagger}_{i,r}$ ($\hat{d}^{\sigma}_{i,r}$) is the creation (annihilation) operator for the orbital $i$ and the site $r$.
$r=0$ denotes the impurity site, and hence $\hat{d}^{\sigma}_{i,0} = \hat{c}_{i\sigma}$ and $\tilde{\epsilon}_{i0} = -\tilde{\mu}$. 
$r \ge 1$ denotes an infinite chain of the bath sites attached to the impurity site. 
With a proper choice of the gauge, all the hopping parameters $t_{ir}$ can be taken to be non-negative, i.e., $t_{ir} \ge 0$. 
The weight for a configuration $C_n$ is 
\begin{eqnarray}\label{Eq_werner_wt}
W(C_n) = {\rm Tr} \! \! \! \! &\bigl[ &  \! \! e^{-(\beta-\tau_n)\tilde{\mathcal H}_0 }  V(\kappa_n, s_n)  
   e^{-(\tau_n  - \tau_{n-1})  \tilde{\mathcal H}_0}   \nn &\times& \!
     V{( \kappa_{n-1}, s_{n-1})}   \cdots e^{-\tau_1\tilde{\mathcal H}_0} \bigr ],  
\end{eqnarray}
where $V(\kappa_p, s_p)$ represents a vertex of the type $\kappa_p$ and of the auxiliary spin $s_p$, which is inserted at the imaginary time $\tau_p$: For example, for one of the spin-flip terms ($\kappa_p=7$) with $\delta_3=0$, it is written as 
\begin{eqnarray}
V(\kappa_p, s_p) = -\frac{J_{\rm H}d\tau }{4}  \hat{c}^{\dagger}_{1\uparrow}(\tau_p) \hat{c}^{\ }_{2\uparrow} (\tau_p) 
 \hat{c}^{\dagger}_{2\downarrow}(\tau_p) \hat{c}^{\ }_{1\downarrow}(\tau_p).  
\end{eqnarray}
On the chain basis, it has been shown that all the elements of the $e^{-\tau \tilde{\mathcal H}_0}$ matrix are non-negative, which is also true for the density-type vertices $V(\kappa, s)$ irrespective to the spin orientation $s=\pm1$.~\cite{0305-4470-38-48-004,Werner_private_comm}
On the other hand, for the non-density-type vertex, it is easy to see that all the elements of the $-V(\kappa, s)$ matrix are non-negative. 
Since the non-density-type vertices always appear in pair within the double-vertex update, the product involving the pair of the vertices $V(\kappa, s)$ is always non-negative.
Then, the weight $W(C_n)$ turns out to be the trace of the product of the matrices with non-negative elements, and therefore it is non-negative. Although we need a finite $\delta_3$ for the submatrix update, a similar pair cancellation of the negative factors of the vertices will work as far as  $\delta_3$ is small.

\bibliographystyle{apsrev}
\bibliography{resub}
\end{document}